\documentclass[lettersize,journal]{IEEEtran}
\usepackage{amsmath,amsfonts}
\usepackage{algorithmic}
\usepackage{algorithm}
\usepackage{array}
\usepackage{textcomp}
\usepackage{stfloats}
\usepackage{url}
\usepackage{verbatim}
\usepackage{graphicx}
\usepackage{amssymb} 
\usepackage{amsfonts}
\usepackage{amsmath} 
\usepackage{physics} 
\usepackage{algorithmic}  
\usepackage{algorithm} 
\usepackage{graphicx}
\usepackage{color} 
\usepackage{bm}
\usepackage{epstopdf} 
\usepackage{subfigure}
\usepackage{booktabs}  
\usepackage[dvipsnames]{xcolor} 
\usepackage{cite} 
\usepackage{balance}
\usepackage{setspace}  
\usepackage{amsthm}
\usepackage{amsmath} 
\usepackage{hyperref}
\usepackage{balance}
\usepackage{placeins} 
\usepackage{scalerel}
\usepackage{caption}

\allowdisplaybreaks 

\newtheorem{ppn}{Proposition}
 
\newtheorem{remark}{Remark}

\newtheorem*{pf}{Proof}  

\newcommand{\ppnref}[1]{\textbf{Proposition \ref{#1}}}

\newcommand{\algref}[1]{\textbf{Algorithm \ref{#1}}}

\newcommand{\figref}[1]{Fig. \ref{#1}}
\newcommand{\secref}[1]{Section \ref{#1}}

\definecolor{softred}{RGB}{200, 50, 50}

\DeclareMathAlphabet{\mathsfit}{\encodingdefault}{\sfdefault}{m}{sl}
\SetMathAlphabet{\mathsfit}{bold}{\encodingdefault}{\sfdefault}{bx}{n}
\newcommand{\tens}[1]{\bm{\mathsfit{#1}}}
\def\tA{{\tens{A}}}

\def\tD{{\tens{D}}}

\def\tF{{\tens{F}}}

\def\tM{{\tens{M}}}

\def\tP{{\tens{P}}}

\def\tW{{\tens{W}}}
\def\tX{{\tens{X}}}
\def\tY{{\tens{Y}}}

\begin{document}

\bstctlcite{IEEEexample:BSTcontrol}
	
\title{Statistical CSI-Based Distributed Precoding Design for OFDM-Cooperative Multi-Satellite Systems}
	
\author{Yafei Wang, \textit{Graduate Student Member}, \textit{IEEE},  Vu Nguyen Ha, \textit{Senior Member}, \textit{IEEE}, \\Konstantinos Ntontin, \textit{Member}, \textit{IEEE}, Hong Yan, Wenjin Wang, \textit{Member}, \textit{IEEE}, \\Symeon Chatzinotas, \textit{Fellow}, \textit{IEEE}, Björn Ottersten, \textit{Fellow}, \textit{IEEE}
    \thanks{Manuscript received xxx.}
    \thanks{Yafei Wang and Wenjin Wang are with the National Mobile Communications Research Laboratory, Southeast University, Nanjing 210096, China, and also with Purple Mountain Laboratories, Nanjing 211100, China (E-mail: \{wangyf, wangwj\}@seu.edu.cn).}
    \thanks{Vu Nguyen Ha, Konstantinos Ntontin, Symeon Chatzinotas and Björn Ottersten are with the Interdisciplinary Centre for Security, Reliability and Trust (SnT), University of Luxembourg (E-mails: \{vu-nguyen.ha, kostantinos.ntontin, symeon.chatzinotas, bjorn.ottersten\}@uni.lu).}
    \thanks{Hong Yan is with Shanghai Satellite Network Research Institute Co., Ltd., Shanghai 200120, China, Shanghai Key Laboratory of Satellite Network, Shanghai 200120, China, and also with State Key Laboratory of Satellite Network, Shanghai 200120, China (howardyan@139.com).}}
		
    \markboth{}%
    {Shell \MakeLowercase{\textit{et al.}}: A Sample Article Using IEEEtran.cls for IEEE Journals}
    
    \maketitle

	\begin{abstract}
This paper investigates the design of distributed precoding for multi-satellite massive MIMO transmissions. We first conduct a detailed analysis of the transceiver model, in which delay and Doppler precompensation is introduced to ensure coherent transmission. In this analysis, we examine the impact of precompensation errors on the transmission model, emphasize the near-independence of inter-satellite interference, and ultimately derive the received signal model. Based on such signal model, we formulate an approximate expected rate maximization problem that considers both statistical channel state information (sCSI) and compensation errors. Unlike conventional approaches that recast such problems as weighted minimum mean square error (WMMSE) minimization, we demonstrate that this transformation fails to maintain equivalence in the considered scenario. To address this, we introduce an equivalent covariance decomposition-based WMMSE (CDWMMSE) formulation derived based on channel covariance matrix decomposition. Taking advantage of the channel characteristics, we develop a low-complexity decomposition method and propose an optimization algorithm. 
    To further reduce computational complexity, we introduce a model-driven 
    scalable deep learning (DL) approach that leverages the equivariance of the mapping from sCSI to the unknown variables in the optimal closed-form solution, enhancing performance through novel dense Transformer network and scaling-invariant loss function design. Simulation results validate the effectiveness and robustness of the proposed method in some practical scenarios. We also demonstrate that the DL approach can adapt to dynamic settings with varying numbers of users and satellites.
\end{abstract}
	

	%
	\IEEEpeerreviewmaketitle
	
\vspace{-3mm}	
	
	\section{Introduction}

\vspace{-1mm}
    
	\IEEEPARstart{D}{espite} the remarkable advancements in terrestrial communication networks, the coverage of cellular and broadband internet access remains insufficient worldwide \cite{Ericsson2024}. 
    As the sixth-generation (6G) of wireless networks approaches, satellite communication (SatCom) emerges as a pivotal technology, promising truly ubiquitous connectivity, one of the cornerstone vision defined by IMT-2030 
    \cite{10820534}. 
    Satellite mobile networks, particularly those supporting handheld devices, inherently aim to enable global broadband access, overcoming the limitation of terrestrial infrastructure \cite{3GPP_TR_38_811, 3gpp_tr_38_821}. 
    To meet the stringent performance requirements of 6G systems, particularly in terms of spectral efficiency and user experience, precoding has emerged as a critical enabling technology. Using advance antenna arrays and by managing spatial resources and steering multiple beams toward different users, precoding provides sufficient gains to target handhelds and allows data streams to coexist within the same time-frequency resource block, thus enhancing spatial multiplexing and overall system coverage and throughput \cite{9852737}. However, the application of precoding in SatCom systems introduces several unique challenges: long propagation delays, high Doppler shifts, and the difficulty of acquiring instantaneous channel state information (CSI), all stemming from the dynamic and mobile nature of satellite-user propagation geometry \cite{DeFilippo2024CellFree}.
    
    Given these constraints, utilizing statistical CSI (sCSI), estimated by stable auxiliary satellite position data such as ephemeris data and angle of departure (AoD), emerges as a practical and robust solution. Exploiting sCSI not only reduces signaling overhead and feedback requirements but also provides robust performance in fast-varying environments typical of low Earth orbit (LEO) satellite networks. Within this context, a well-designed precoding strategy is fundamental to unlocking the full potential of 6G SatCom systems. It directly contributes to key performance indicators such as signal-to-interference-plus-noise ratio (SINR), signal-to-leakage-plus-noise ratio (SLNR), and achievable sum rate—metrics that underpin the envisioned enhanced user experience, energy efficiency, and coverage gains targeted by next-generation air interface designs. This paper addresses the critical problem of distributed precoding for OFDM-based SatCom under sCSI, offering novel solutions aligned with the performance goals and standardization directions of 6G and beyond.
\vspace{-3mm}
    \subsection{Related Works}   
   Prior research has extensively investigated various precoding techniques for single-satellite systems under practical impairments and constraints. Authors in \cite{8353925} investigated minimum mean square error (MMSE)-based precoding, scheduling, and link adaptation techniques and analyzed the impact of outdated CSI. The robustness against phase errors and CSI impairments has also been addressed through power-minimization precoding in \cite{8629918} and resource efficiency maximization approaches in \cite{wang2021resource}.  
    Enhancing satellite precoding by scaling the antenna array size is another critical research direction. In particular, \cite{you2020massive} conducted a characteristic analysis of massive multi-input multi-output (MIMO) channels and designed sCSI-based precoding solutions to improve the average SLNR performance.
    Furthermore, recognizing the degradation introduced by outdated precoding vectors and dynamic satellite environments, \cite{10437228} developed a low-complexity precoding update algorithm that effectively mitigates performance losses due to temporal CSI misalignment.

   Recently, driven by the need for intelligent, adaptive, and low-complexity solutions for limited CSI availability or high-dimensional optimization problems, 
   deep learning (DL) techniques have been studied for SatCom precoding designs (PDs). For instance, \cite{9815078} constructed a convolutional neural network (CNN) that directly learns the key matrices, significantly reducing the computational overhead associated with conventional power-minimization-aimed precoding methods.  
   Exploiting the deep reinforcement learning (DRL) techniques, \cite{10336551} developed a hybrid precoding algorithms, leveraging Conditional Value at Risk (CVaR) to maximize energy efficiency and ensure QoS guarantees under dynamic and uncertain channel conditions.
   In \cite{10550141}, a joint channel prediction and PD was studied by using long short-term memory and variational autoencoder, which are jointly trained to anticipate channel evolution and optimize precoding decisions accordingly. 

   Due to current industrial limitations in satellite payload and antenna manufacturing, the communication capacity of a single satellite, bounded by its link budget, still faces limitations. 
   To overcome this and meet the rising service demands of 6G, a promising solution is to deploy dense LEO satellite constellations and leverage inter-satellite cooperative transmission techniques, enabling distributed MIMO and cell-free architectures. 
   In particular, \cite{9939157} has introduced a cell-free LEO satellite scheme and proposed two joint power allocation and handover schemes that mitigate inter-satellite interference and improve user experience continuity efficiently. 
   Further deepening the theoretical understanding, \cite{10061620} derived closed-form expressions for spectral efficiency in LEO-based distributed MIMO systems, offering a comparative analysis between single- and multi-satellite transmission scenarios. While \cite{10380500} explored hybrid precoding architectures in power- and cost-constrained multiple-satellite systems.    
    As another form of cooperative transmission, the terminal-side spatial multiplexing, where terminals with multiple antennas receive different data streams from satellites, is studied in \cite{10440321}.
    With inter-satellite cooperation, \cite{ha2024UCB} investigates a low-complexity transmission paradigm on beam domain, where user-beam pairing and precoding are designed based on earth-moving beamforming to improve the SINR. 
    Focusing on the impact of asynchrony on coherent transmission, \cite{10596023} and \cite{10615897} model the errors introduced by asynchrony based on perfect instantaneous CSI, construct asynchronous signal models under the DVB-S2X standard, and design an asynchronous weighted MMSE (WMMSE) algorithm along with a delay estimation algorithm. Recently, \cite{wu2025distributed} analyzes the effect of synchronization errors on the performance of distributed beamforming in dual-satellite OFDM systems, but lacks exploration of error distribution modeling and rate-maximizing transmission.

\vspace{-3mm}
\subsection{Contributions}
    To enable direct communication with unmodified terrestrial handheld terminals using terrestrial standard \cite{3GPP_TR_38_811, 3gpp_tr_38_821} and to uphold ubiquitous connectivity in the 6G era, it is essential to address the challenges of delay and Doppler compensation in fast-moving LEO SatCom environment, along wtih distributed precoding in OFDM systems, which have emerged as critical yet underexplored research directions. 
    These challenges are exacerbated by the inherent difficulty of acquiring accurate and instantaneous CSI, motivating the need for robust and practical PDs that can operate effectively with only sCSI.
    
    While recent efforts have made strides in closed-form and heuristic PDs, such approaches often fall short of revealing or approaching the theoretical performance limits, especially in complex, distributed satellite networks. This gap highlights a fundamental and timely research question:
    \textit{How can we design efficient and scalable distributed precoding schemes for OFDM-based SatCom systems under sCSI, capable of meeting the stringent demands of 6G connectivity?}
    This paper addresses this question by introducing a novel framework for distributed PD in LEO satellite networks, specifically designed to overcome the limitations of existing methods and to advance the state of the art in sCSI-based precoding. The major contributions of this work are as follows:
 \begin{itemize}
    \item We conduct a detailed analysis of the transceiver processing flow in OFDM-based satellite distributed MIMO networks. Within this process, we investigate the various impacts caused by delay and Doppler compensation errors, with a emphasis on the high incoherence of inter-satellite interference signal. Ultimately, we derive the received signal model for satellite distributed MIMO under compensation errors. By further considering the challenges in acquiring iCSI, we formulate an approximate expected sum-rate maximization problem for the design of satellite distributed precoding with sCSI.
    \item Based on the formulated problem and existing literature, we first reformulate it into a WMMSE problem. However, we prove that this formulation is not equivalent to the original problem and analyze the conditions for equivalence, thereby exposing its limitations. Accordingly, we construct a covariance decomposition-based WMMSE (CDWMMSE) problem whose objective function, derived from channel covariance matrix decomposition, is equivalent to the original. Exploiting the LoS-dominant nature of satellite channels, we propose a low-complexity decomposition method and develop an alternating optimization-based solution algorithm.
    \item 
    To reduce the computational complexity, we propose a model-driven 3D scalable DL method. We first construct a mapping from sCSI to the unknown variables in the optimal closed-form solution and prove the equivariance and invariance it satisfies. To approximate this mapping, we propose the dense Transformer network (DTN). By analyzing the relationship between key variables in the optimal closed-form precoding expression and the sum rate, we propose a more compatible scaling-invariant loss function, which facilitates efficient training under supervised learning. Owing to these designs, the network approaches optimization-based performance with reduced complexity while supporting 3D scalability, enabling training under a fixed configuration and deployment in dynamic scenarios with varying numbers of users, satellites, and transmit antennas.
\end{itemize}

This paper is structured as follows: The system model and signal model are built in \secref{signal model sec}. \secref{CDWMMSE sec} formulates CDWMMSE problem and the algorithm. \secref{modified WMMSE DL sec} proposes the DL-based algorithm. \secref{Section 6} reports the simulation results, and the paper is concluded in \secref{conclusion sec}.

	{\textit{Notation}}: $x, {\bf x}, {\bf X}$, and $\tX$ represent scalar, column vector, matrix, and tensor. $(\cdot)^T$, $(\cdot)^{*}$, $(\cdot)^H$, and $(\cdot)^{-1}$ denote the transpose, conjugate, transpose-conjugate, and inverse operations, respectively. ${\bf I}_{M}$ represents $M\times M$ identity matrix. $\left \|\cdot\right \|_{2}$ denotes $\ell_2$-norm. $\otimes$ and $\odot$ are the Kronecker product and Hardmard product operations. The operator ${\rm \textbf{Tr}}(\cdot)$ represents the matrix trace.  ${{\rm diag}\{{\bf a}\}}$ represents a diagonal matrix whose diagonal elements are composed of ${\bf a}$. $\real(\cdot)$ and $\imaginary(\cdot)$ denote the real and imaginary parts of a complex scalar, vector, or matrix. We use $\tX_{[m_1,...,m_N]}$ to denote the indexing of elements in $\tX\in \mathbb{R}^{M_1\times \cdots\times M_N}$. $[\tA_1,...,\tA_K]_{S}$ denotes the tensor formed by stacking $\tA_1,...,\tA_K$ along the $S$-th dimension. ${\bf x}{\succeq}{\bf 0}$ means all the elements of ${\bf x}$ is nonnegative. The expression $\mathcal{C}\mathcal{N}(\mu, \sigma^2)$ denotes circularly symmetric Gaussian distribution with expectation $\mu$ and variance $\sigma^2$. ${\mathbb{R}}^{M\times N}$ and ${\mathbb{C}}^{M\times N}$ represent the set of $M\times N$ dimension real- and complex-valued matrixes. $\nabla f$ denotes gradient of function $f(\cdot)$. $k\in \mathcal{K}$ means element $k$ belongs to set $\mathcal{K}$.

    \vspace{-2mm}
    \section{mSatCom Distributed Precoding System Model}
    \label{signal model sec}
    \vspace{-1mm}

    \begin{figure}[!t]
		\centering
		\includegraphics[width=0.4\textwidth, height = 5cm]{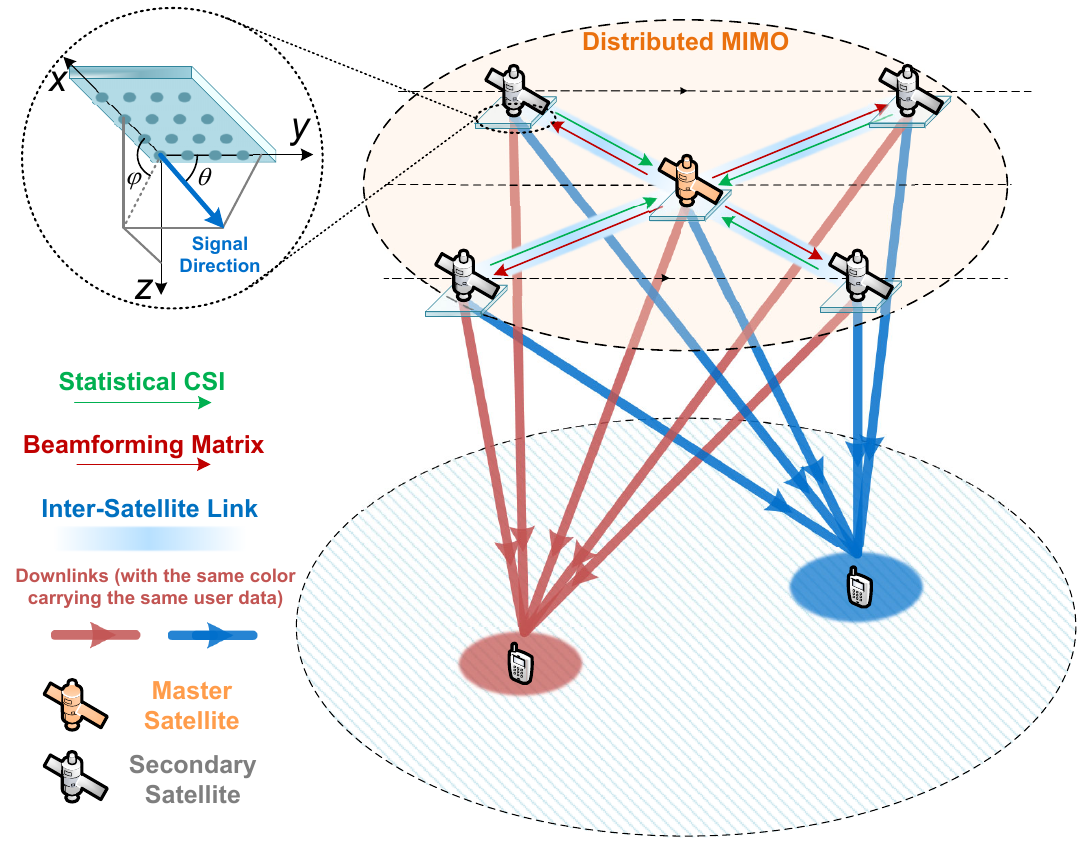}
		\caption{mSatCom cooperative distributed precoding system.}
            \label{system framework}
            \vspace{-5mm}
	\end{figure}

    We consider the downlink cooperative transmission of a multi-satellite communication (mSatCom) system, where \( S \) satellites simultaneously provide service to \( K \) user terminals (UTs) within the same time-frequency resources. Each satellite is equipped with a uniform planar array (UPA) comprising \( N_{\rm T} = N_{\rm v} N_{\rm h} \) antennas, whereas each UT is assumed to have a single antenna. 
    Assume that one of these satellites is selected as the \textit{``master''} to coordinate cooperative transmission \cite{3gpp_tr_38_821} according to the service quality metric, typically close to the region center, as shown in \figref{system framework}. 
    These other satellites are so-called ``\textit{secondary}'' ones.
    In this cooperative transmission architecture, coherent transmission is achieved by precompensation enabling distributed precoding across satellites to deliver identical data streams to the UTs. Without loss of generality, we assume that user data streams are sent from the ground gateway to the master satellite, which then relays them to the secondary satellites.

\vspace{-3mm}
    \subsection{Channel Model} \label{channel model sec}
	The base-band time-frequency-spatial channel ${\bf h}_{s,k}(t,f)\in\mathbb{C}^{N_{\rm T}\times 1}$ from the $s$-th satellite to the $k$-th UT is \cite{you2020massive}
	\begin{align}
		\begin{split}
        &{\bf h}_{s,k}(t,f)\!=\! a_{s,k}(t,f){\rm e}^{j2\pi(t\nu_{s,k}^{\rm sat}-f\tau_{s,k}^{\rm min})}{\bf v}({\boldsymbol{\theta}}_{s,k}),
		\end{split}
	\end{align}
   The term \( a_{s,k}(t,f) \) represents the complex channel coefficient comprising both line-of-sight (LoS) and non-line-of-sight (NLoS) components. It is modeled as a Rician fading variable with Rician factor \( \kappa_{s,k} \) and an average power given by \( \mathbb{E}_{t,f}\big[|a_{s,k}(t,f)|^2\big] = \gamma_{s,k} \) \cite{you2020massive, 3GPP_TR_38_811, 8840846}. The term \( \exp\{j2\pi(t\nu_{s,k}^{\rm sat} - f\tau_{s,k}^{\rm min})\} \) captures the frequency- and time-dependent phase rotation, where \( \nu_{s,k}^{\rm sat} \) denotes the Doppler shift resulting from satellite motion, and \( \tau_{s,k}^{\rm min} \) is the propagation delay associated with the LoS path. The vector \( \mathbf{v}(\boldsymbol{\theta}_{s,k}) = \mathbf{v}(\theta^{\rm x}_{s,k}, \theta^{\rm y}_{s,k}) \) characterizes the angular relationship between the antenna array plane and the UT, and is constructed based on the azimuth and elevation angles of departure from satellite \( s \) to user \( k \), expressed as
    \begin{align}
        {\bf v}({\boldsymbol{\theta}}_{s,k}) &= {\bf v}_{N_{\rm v}}(\cos (\smash{\theta^{\rm y}_{s,k}}))\otimes  {\bf v}_{N_{\rm h}}(\sin(\smash{\theta^{\rm y}_{s,k}})\cos(\theta^{\rm x}_{s,k})),\\
        {\bf v}_{N}(x) &= (1/\sqrt{N_{\rm T}}) [{\rm e}^{-j\pi0x}, {\rm e}^{-j\pi1x}, ..., {\rm e}^{-j\pi(N-1)x}].
    \end{align}
    
\vspace{-5mm}

    \subsection{Transmitter Delay and Doppler Precompensation}
   In order to enable coherent transmission across multiple satellites and guarantee correct OFDM demodulation at the receiver side, the OFDM modulation process must be carried out separately for each user at the transmitter. The entire system bandwidth is utilized for OFDM transmission, where all users occupy the full set of subcarriers, and user scheduling is achieved through power allocation. After applying the inverse Fourier transform (IFT), the time-domain baseband signal corresponding to symbols $0, \ldots, M\!-\!1$, transmitted from satellite $s$ to UT $k$, is expressed as
    \begin{align}
        {\bf x}_{s,k}(t)\! = \! \scaleobj{.8}{\sum_{n=0}^{N-1}\sum_{m=0}^{M-1}}\!{\bf x}^{(m)}_{s,k,n} u(t\!-\!mT_{\rm sym}){\rm e}^{j2\pi n\Delta f (t-mT_{\rm sym})},
	\end{align}
    where $u(t) = 1$ if $t\in[0, T]$ and $u(t) = 0$ otherwise; $T_{\rm sym} = T_{\rm cp}+T$, and ${\bf x}^{(m)}_{s,k,n}$ is transmit signal at the $n$-th sampling of $m$-th OFDM symbol. After appending the cyclic prefix (CP) and applying frequency upconversion,\footnote{In this work, upconversion is assumed to be performed prior to delay compensation. In practical implementations, the order of these operations may vary; however, this does not affect the validity of the subsequent analysis.} the transmit signal can be written as
\begin{align}
    \hspace{-3mm}{\hat {\bf x}}_{s,k}(t) \!=\! 
    \begin{cases}
     {\rm e}^{j2\pi f_0 t}{\bf x}_{s,k}(t\!+\!\!T),\ \!  t\!\in\![mT_{\rm sym}\!\!-\!T_{\rm cp}, mT_{\rm sym}),\\
    {\rm e}^{j2\pi f_0 t}{\bf x}_{s,k}(t),\ \  t\!\in\![mT_{\rm sym}, mT_{\rm sym}\!+\!T),
    \end{cases}\!\!\!\!\!\!\!
\end{align}
To achieve coherent transmission, unlike the commonly used receiver-side synchronization \cite{8795582}, delay and Doppler precompensation must be performed at the transmitter. The compensated signal is given by:
\begin{align}
        {\hat {\bf x}}^{{\rm cps}}_{s,k}(t) = {\hat {\bf x}}_{s,k}(t+\tau_{s,k}^{\rm cps}){\rm e}^{-j2\pi (t+\tau_{s,k}^{\rm cps})\nu_{s,k}^{\rm cps}},
        \label{precompensation}
    \end{align}
    where \( \tau_{s,k}^{\rm cps} \) and \( \nu_{s,k}^{\rm cps} \) denote the delay and Doppler compensation applied by the satellite $s$ for UT $k$. In practical satellite deployments, Doppler shift and delay can be estimated at the UT based on downlink pilot signals \cite{8795582}, with predictions employed to against parameter time variation. These estimated parameters are then fed back to the satellite for precompensation.
    The corresponding bandpass received signal (``without noise'') is obtained through the convolution between the time-domain transmit signal and the channel impulse response, which accounts for both delay and Doppler effects, i.e.,
    \vspace{-1mm}
\begin{align}
    {\bar y}_{s,k}(t) 
    &= \scaleobj{.8}{\sum_{l=1}^{K}\sum_{\forall s}\int}  {\bf h}^T_{s,k}(t,\tau){\hat {\bf x}}^{{\rm cps}}_{s,l}(t-\tau){\rm d}\tau\notag\\
    &= \scaleobj{.8}{\sum_{l=1}^{K}\sum_{\forall s}} {\breve {\bf h}}^T_{s,k}(t,\tau_{s,k}^{\rm min}) {\hat {\bf x}}_{s,l}(t+\tau_{s,l}^{\rm cps}-\tau_{s,k}^{\rm min})\notag\\
    &\qquad \times {\rm e}^{j2\pi t(\nu^{\rm sat}_{s,k}-\nu_{l,k}^{\rm cps})} {\rm e}^{-j2\pi (\tau_{s,l}^{\rm cps}-\tau_{s,k}^{\rm min})\nu_{s,l}^{\rm cps}},
    \vspace{-1mm}
\end{align}
    where ${\breve {\bf h}}^T_{s,k} = a_{s,k}(t, \tau_{s,k}^{\rm min}){\bf v}^T({\boldsymbol{\theta}}_{s,k})$. 
    We define the compensation error for UT $k$ and the impact of the compensation for UT $l$'s signal on the interference to UT $k$ as follows:
    \begin{align}
        {\bar \tau}_{s,k} = \tau_{s,k}^{\rm cps}-\tau_{s,k}^{\rm min},&\ {\bar \nu}_{s,k} = \nu^{\rm sat}_{s,k}-\nu_{s,k}^{\rm cps},\\
        {\tilde \tau}_{s,l,k} = \tau_{s,l}^{\rm cps}-\tau_{s,k}^{\rm min},&\ {\tilde \nu}_{s,l,k} = \nu^{\rm sat}_{s,k}-\nu_{s,l}^{\rm cps},
    \end{align}
   where the errors ${\bar \tau}_{s,k}$ and ${\bar \nu}_{s,k}$ originate from UT-side estimation inaccuracies and feedback-induced imperfections, and their distributions depend on the specific estimation and compensation algorithms employed. At the receiver, the time-domain baseband signal of the $m$-th OFDM symbol is ${\bar y}^{(m)}_{s,k}(t) = {\bar y}_{s,k}(t){\hat u}(t-mT_{\rm sym})$, where ${\hat u}(t) = 1$  if $t\in[-T_{\rm cp}, T]$ and ${\hat u}(t) = 0$, otherwise. In the following, we individually examine the desired signal component and the interference terms.

   \vspace{-3mm}
    \subsection{Desired Received Signal Model}
    \vspace{-1mm}
        
   The desired signal component received by the $k$-th user from the $s$-th satellite is expressed as
    \begin{align}
        {\breve{\bf h}}^T_{s,k}(t,\tau_{s,k}^{\rm min}) {\hat {\bf x}}_{s,k}(t\!+\!{\bar \tau}_{s,k}) {\rm e}^{j2\pi (t{\bar \nu}_{s,k}-{\bar \tau}_{s,k}\nu_{s,k}^{\rm cps})},\  s\in\mathcal{S}.
    \end{align}
    We denote the delay compensation error as ${\bar \tau}_{s,k}=N^{\rm de}_{s,k}T_s+{\hat \tau}_{s,k}$, where $|{\hat \tau}_{s,k}|<T_s$ and ${\hat \tau}_{s,k}$ has the same sign as $N^{\rm de}_{s,k}$. Regarding ${\bar \tau}_{s,k}$, three distinguished cases can be considered as: (i) ${\bar \tau}_{s,k}<0$ with $|{\bar \tau}_{s,k}| < T_{\rm cp}$; (ii) ${\bar \tau}_{s,k}<0$ with $|{\bar \tau}_{s,k}| > T_{\rm cp}$; and (iii) ${\bar \tau}_{s,k}>0$. Among these, only the first case—where the sampling window starts within the CP duration—does not introduce inter-symbol interference (ISI). We assume the first case, where synchronization can at least achieve symbol-level signal alignment \cite{8795582}. Under this assumption, the downconversion and sampling results for the effective signal in this case are given as
    \begin{align}
        {\hat {\bf x}}^{(m)}_{s,k}(i) \!=\! 
        \begin{cases}
        \sum_{n=0}^{N-1}{\bf x}^{(m-1)}_{s,k,n}\varpi^{s,k}_{n,i},\  i \!=\! 0,...,N^{\rm de}_{s,k}\!\!-\!\!1\\
        \sum_{n=0}^{N-1}{\bf x}^{(m)}_{s,k,n}\varpi^{s,k}_{n,i}, \  i \!= \!N^{\rm de}_{s,k},...,N_{\rm sym}\!\!-\!\!1
        \end{cases}\!\!\!,
    \end{align}
    where $\varpi^{s,k}_{n,i} = {\rm e}^{j2\pi \{ f_0{\bar \tau}_{s,k} + n [\frac{i}{N}+\Delta f({\bar \tau}_{s,k}-T_{\rm cp})]\}}$.
    In this case, $|{\bar \tau}_{s,k}| =N^{\rm de}_{s,k}T_s+{\hat \tau}_{s,k} < T_{\rm cp}=N_{\rm cp}T_s$ leads to $N^{\rm de}_{s,k} <N_{\rm cp}$, thereby ensuring that the signal component ${\bf x}^{(m-1)}_{s,k,n}$ from the previous OFDM symbol is entirely removed during CP elimination. It is further assumed that the wireless channel remains static throughout a single OFDM symbol duration. Following downconversion, sampling, and CP removal, the resulting received signal can be expressed as
    \vspace{-1mm}
    \begin{align}
        {\bar y}^{{\rm eff}(m)}_{k}(i) = \scaleobj{.8}{\sum_{\forall s}}  {\breve{\bf h}}^T_{s,k}  \scaleobj{.8}{\sum_{n=0}^{N-1}}{\bf x}^{(m)}_{s,k,n} {\bar \varpi}^{s,k}_{n,i},
        \ \  i \in \mathcal{N},
        \vspace{-1mm}
    \end{align}
    where $\mathcal{N} = \{0,...,N-1\}$ and 
    \begin{align}
        &{\bar \varpi}^{s,k}_{n,i} = \varphi^n_{s,k}{\rm e}^{j2\pi [n(\frac{i}{N}+\Delta f{\bar \tau}_{s,k})+ iT_s(\nu^{\rm sat}_{s,k}-\nu_{s,k}^{\rm cps})]},\\
        &\varphi^n_{s,k} = {\rm e}^{j2\pi (f_0+n\Delta f-\nu_{s,k}^{\rm cps}){\bar \tau}_{s,k}}.
        \label{phase error}
    \end{align}
    Applying the DFT, the frequency-domain representation of the desired received signal is obtained as
    \vspace{-1mm}
    \begin{align}
        &{\bar y}^{{\rm eff}(m)}_{k,j} \!\!\! = \! \scaleobj{.8}{\sum_{s=1}^{S}}{\breve{\bf h}}^T_{s,k}  \scaleobj{.8}{\sum_{n=0}^{N-1}}\varphi^n_{s,k}{\bf x}^{(m)}_{s,k,n}\psi({\bar \nu}_{s,k}, n\!-\!j),\ j \!\in\! \mathcal{N},\\
        &\psi({\bar \nu}_{s,k}, n-j) =\frac{1}{N}\frac{1-{\rm e}^{j2\pi \left[{\bar \nu}_{s,k}/{\Delta f}-(n-j)\right]}}{1-{\rm e}^{j2\pi \left[{\bar \nu}_{s,k}/{\Delta f}-(n-j)\right]/N}}.
        \vspace{-1mm}
    \end{align}
    The function $\psi(\cdot)$ captures the inter-carrier interference (ICI) resulting from the Doppler estimation error \( {\bar \nu}_{s,k} \). It is assumed that the residual frequency offset after compensation is significantly smaller than the subcarrier spacing, i.e., \( \nu^{\rm sat}_{s,k} - \nu_{s,k}^{\rm cps} \ll \Delta f \). Under this assumption, the desired received signal at $k$-th UT on the $n$-th subcarrier can be formulated as
    \begin{align}
         {\bar y}^{{\rm eff}(m)}_{k,n} \approx \scaleobj{.8}{\sum_{\forall s}}{\breve{\bf h}}^T_{s,k} \ {\bf x}^{(m)}_{s,k,n}\varphi^n_{s,k}.
    \end{align}
    
    \vspace{-5mm}
        
    \subsection{Interfering Received Signal Model}\label{Interference sec}
    The interference signal from the $s$-th satellite and the $l$-th ($l\neq k$) user is $ {\breve{\bf h}}^T_{s,k}{\hat {\bf x}}_{s,l}(t+{\tilde \tau}_{s,l,k})\cdot {\rm e}^{j2\pi (t{\tilde \nu}_{s,k}-{\tilde \tau}_{s,k}\nu_{s,k}^{\rm cps})}$.
    To simplify the subsequent formulations, the interference signal at $k$-th UT on the $n$-th subcarrier after downconversion, sampling, CP removal, and DFT is denoted as
    \vspace{-1mm}
    \begin{align}
        \textstyle{\bar y}^{{\rm interf}(m)}_{k,n} \approx \scaleobj{.8}{\sum_{\forall s}\sum_{l=1,l\neq k}^{K}}{\breve{\bf h}}^T_{s,k} \ {\breve {\bf x}}^{(m'_{s,l,k})}_{s,l,n}\varphi^n_{s,l,k}.
        \vspace{-1mm}
    \end{align}
    
\begin{figure}[t]
        \centering
        \includegraphics[width=0.45\textwidth]{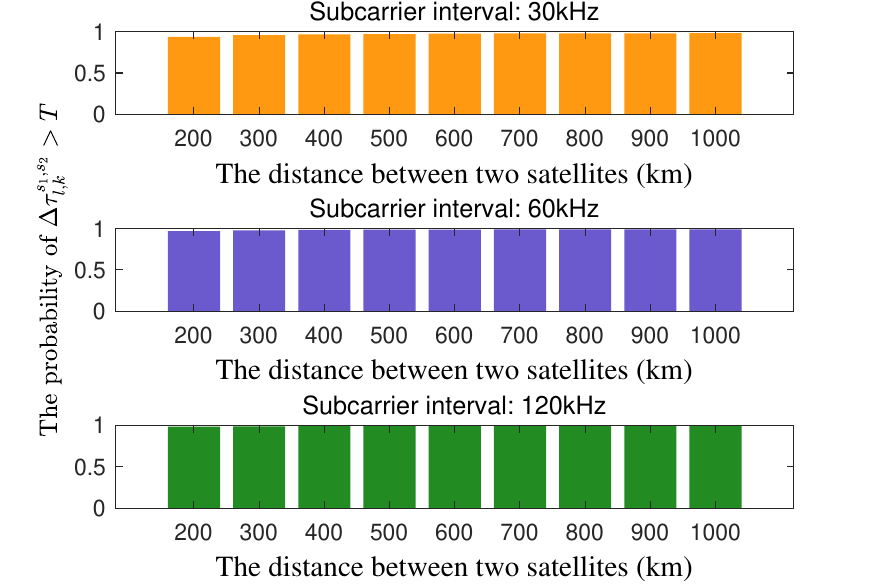}
        \vspace{-2mm}
        \caption{Probability that inter-satellite interference signals are statistically independent.}
        \label{Symbol Index fig}
        \vspace{-4mm}
    \end{figure}

   \begin{remark} 
    The interference signal is characterized by the following key features: 
    \begin{enumerate}
        \item \textbf{ISI\&ICI:} Due to the potentially large values of \( |{\tilde \tau}_{s,l,k}| \) and \( |{\tilde \nu}_{s,l,k}| \), \( {\breve {\bf x}}^{(m'_{s,l,k})}_{s,l,n} \) may encompass contributions from multiple OFDM symbols and several subcarriers, namely \( {\bf x}^{(m'_{s,l,k})}_{s,l,1}, \dots, {\bf x}^{(m'_{s,l,k})}_{s,l,N} \), where $m'_{s,l,k}$ denotes the index spanning multiple symbols.
        \item \textbf{Inter-satellite interference independence:} Consider two interference signals at UT $k$ caused by UT $l$'s transmissions from satellites $s_1$ and $s_2$. If the delay difference $\Delta\tau^{s_1, s_2}_{l,k} = |(\tau_{s_1,l}^{\rm min}-\tau_{s_1,k}^{\rm min}) - (\tau_{s_2,l}^{\rm min}-\tau_{s_2,k}^{\rm min})| \approx |{\tilde \tau}_{s_1,l,k} - {\tilde \tau}_{s_2,l,k}| > T_{\rm sym}$, the interference signals ${\breve {\bf x}}^{(m'_{s_1,l,k})}_{s_1,l,n}$ and ${\breve {\bf x}}^{(m'_{s_2,l,k})}_{s_2,l,n}$ contain symbol information from completely different indexes $m'_{s_1,l,k}$ and $m'_{s_2,l,k}$, and are therefore statistically independent.
        Given the large distances between satellites on the same or different orbital planes, this condition is typically met, as $T_{\rm sym}$ is relatively short \cite{3gpp_tr_38_821}. For satellites at 600~km altitude serving UTs within an 800~km radius, the probability that $\Delta\tau^{s_1, s_2}_{l,k} > T$ is shown in Fig.~\ref{Symbol Index fig}. Since $T_{\rm cp}$ is small, its influence is negligible. Moreover, user scheduling \cite{wu2023low} may further enlarge user spread, increasing this probability.
    \end{enumerate}
    \end{remark}

    Since the PD is based on sCSI, it remains invariant over multiple symbols. For simplicity, the subcarrier index $n$ is omitted. Therefore, the received signal at UT $k$ on the $m$-th OFDM symbol is given by:
    \begin{equation}
        y^{(m)}_{k} \!\!=\!\! \scaleobj{.8}{\sum_{\forall s}}{\breve{\bf h}}^T_{s,k} \! \Big( \! {\bf w}_{s,k}d^{(m)}_{k}\! \varphi_{s,k} \! + \!\! \scaleobj{.8}{\sum_{\forall l\neq k}} {\bf w}_{s,l}{\breve d}^{(m'_{s,l,k})}_{l}\!\varphi_{s,l,k}\! \Big) + n^{(m)}_{k},
    \end{equation}
    where $d^{(m)}_k$ represents the symbol information for the $k$-th UT at $m$-th OFDM symbol, and $n^{(m)}_{k}\sim \mathcal{C}\mathcal{N}(0, \sigma_k^2)$ denotes the additive noise at the $k$-th UT.
    ${\bf w}_{s,k}\in\mathbb{C}^{N_{\rm T}\times 1}$ denotes the precoding vector at satellite $s$ for UT $k$. $\varphi_{s,k}$ with definition \eqref{phase error} is the phase error introduced by delay and Doppler estimation errors. Due to the high carrier frequency $f_0$, completely removing $\varphi_{s,k}$ is practically infeasible, and its statistical distribution is closely tied to the parameter estimation and prediction method mentioned after \eqref{precompensation}.
    
    \vspace{-3mm}    
    \subsection{Sum Rate Expression and Statistical CSI Exploitation}\label{sum rate expression}
        We assume that the information of different UTs is independent, $\mathbb{E}\{d^{(m)}_{k}(d^{(m)}_{k})^H\} = 1$, and $\mathbb{E}\{{\breve d}^{(m'_{s,l,k})}_{l}({\breve d}^{(m'_{s,l,k})}_{l})^H\} = 1$. Due to the independence of inter-satellite interference, we approximate $\mathbb{E}\{{\breve d}^{(m'_{s_1,l,k})}_{s_1,l}({\breve d}^{(m'_{s_2,l,k})}_{s_2,l})^H\} \approx 0$.
    Based on the above analysis, the expected achievable rate is given by 
     \begin{align}
    \mathbb{E}\{R_k\} &\!= \!\mathbb{E}_{\bf H}\left\{\log_2\!\left(\!1\!+\!\scaleobj{.8}{\frac{\sum\nolimits_{s_1=1}^{S}\sum\nolimits_{s_2=1}^{S}\!{\bf w}^H_{s_1,k}{\bar {\bf h}}^*_{s_1,k}{\bar {\bf h}}^T_{s_2,k}{\bf w}_{s_2,k}}{\sum_{i\neq k}\sum_{s=1}^{S}{\bf w}^H_{s,i}{\bar {\bf h}}^*_{s,k}{\bar {\bf h}}^T_{s,k}{\bf w}_{s,i}+\sigma_k^2}}\!\!\right)\right\}\notag\\
    &\!\approx {\bar R}_k \triangleq \log_2\!\left(\!1\!+\!\scaleobj{.8}{\frac{\sum\nolimits_{s_1=1}^{S}\sum\nolimits_{s_2=1}^{S}{\bf w}^H_{s_1,k}\mathbb{E}\{{\bar {\bf h}}^*_{s_1,k}{\bar {\bf h}}^T_{s_2,k}\}{\bf w}_{s_2,k}}{\sum_{i\neq k}\sum_{s=1}^{S}{\bf w}^H_{s,i}\mathbb{E}\{{\bar {\bf h}}^*_{s,k}{\bar {\bf h}}^T_{s,k}\}{\bf w}_{s,i}+\sigma_k^2}}\right) \notag,
    \vspace{-1mm}
    \end{align}
    where ${\bar {\bf h}}_{s,k} = {\breve {\bf h}}_{s,k}\varphi_{s,k}$. Then, we can simplify ${\bar R}_k$ as
    \vspace{-1mm}
	\begin{align}
		\textstyle{\bar R}_k = \log_2\Big(1+\frac{{\bf w}^H_{k}{\boldsymbol{\Omega}}_k{\bf w}_{k}}{\sum_{i\neq k}{\bf w}^H_{i}{\tilde {\boldsymbol{\Omega}}}_k{\bf w}_{i}+\sigma_k^2}\Big),
		\label{average sumrate}
        \vspace{-2mm}
	\end{align}
    in which 
    \vspace{-2mm}
\begin{align}
    &{\bf w}_k = [{\bf w}^T_{1,k}, {\bf w}^T_{2,k},\cdots, {\bf w}^T_{S,k}]^T\in\mathbb{C}^{SN_{\rm T}\times 1},\\
    &{\bar {\bf h}}_k = [{\bar {\bf h}}^T_{1,k}, {\bar {\bf h}}^T_{2,k},\cdots, {\bar {\bf h}}^T_{S,k}]^H\in\mathbb{C}^{SN_{\rm T}\times 1},\\
    &{\boldsymbol{\Omega}}_{s1|s2,k} = \mathbb{E}\{{\bar {\bf h}}^*_{s_1,k}{\bar {\bf h}}^T_{s_2,k}\},\ {\boldsymbol{\Omega}}_k = \mathbb{E}[{\bar {\bf h}}^*_k{\bar {\bf h}}^T_k],\\
    &{\tilde {\boldsymbol{\Omega}}}_k = {\rm \textbf{BlkDiag}}\{{\boldsymbol{\Omega}}_{1|1,k}, ..., {\boldsymbol{\Omega}}_{S|S,k}\}\in\mathbb{C}^{SN_{\rm T}\times SN_{\rm T}}.
    \vspace{-2mm}
\end{align}
    According to the channel and signal models given in \secref{channel model sec} and \secref{Interference sec}, one yields
    \vspace{-1mm}
    \begin{align}
    &\mathbb{E}\{{\bar {\bf h}}_{s,k}\} = \rho_{s,k}\cdot {\bf v}(\theta^{\rm x}_{s,k}, \theta^{\rm y}_{s,k}), \label{barh eq1}\\
    &\textstyle{\boldsymbol{\Omega}}_{s1|s2,k} \!=\! 
    \begin{cases}
        \gamma_{s,k}{\bf v}^*({\boldsymbol{\theta}}_{s,k}){\bf v}^T({\boldsymbol{\theta}}_{s,k}), \  s_1\!=\!s_2\!=\!s,\\
        \rho_{s_1,k}^*\rho_{s_2,k}  {\bf v}^*({\boldsymbol{\theta}}_{s_1,k}){\bf v}^T({\boldsymbol{\theta}}_{s_2,k}), \  s_1\!\neq\! s_2,
    \end{cases}\\
     &\textstyle\rho_{s,k} = \sqrt{\frac{\kappa_{s,k} \gamma_{s,k}}{2(\kappa_{s,k} + 1)}} (1+j){\bar \varphi}_{s,k},
     \vspace{-3mm}
    \end{align}
    where \(\gamma_{s,k}\), \(\kappa_{s,k}\), \({\boldsymbol{\theta}}_{s,k}\), and \({\bar \varphi}_{s,k} \triangleq \mathbb{E}\{\varphi_{s,k}\}\) denote the average channel power, K-factor, the angle between the transmit antenna array and the UT, and the expected phase error, respectively. These parameters, collectively referred to as the sCSI, are assumed to remain unchanged over an extended period (e.g., several OFDM symbol durations), during which they can be accurately estimated \cite{you2020massive, wang2021resource}.

\vspace{-2mm}
   \section{mSatCom Distributed Precoding Design for \\CDWMMSE Criterion}
   \label{CDWMMSE sec}
    Assuming per-subcarrier power constraints, as commonly adopted in prior work~\cite{yuan2023alternating}, the weighted sum rate (WSR) maximization problem is formulated as
    \begin{align}
                (\mathcal{P}_1):\; \max\limits_{{\bf W}} \; \scaleobj{.8}{\sum_{\forall k}}\beta_k{\bar R}_k \; {\rm s.t.}\ {\rm \textbf{Tr}}\left({\bf W}_s{\bf W}^H_s\right) \leq P_s,\ s\in\mathcal{S},
        \label{WSR problem}
    \end{align}
    where $\beta_k$ denotes the weight assigned to the $k$-th UT, and $P_s$ represents the maximum allowable transmit power of the $s$-th satellite on the considered subcarrier. Furthermore, the signal model in Section~\ref{signal model sec} is also compatible with other classical PD objectives, such as those in~\cite{bjornson2014optimal, SB-9931, SB-0105}.

\vspace{-3mm}
\subsection{WSR Criterion vs. WMMSE Criterion}\label{WMMME sec}
Following recent advances in solving WSR problems~\cite{10596023, 10615897}, a conventional approach is to reformulate $(\mathcal{P}_1)$ as a WMMSE problems,
	\begin{align}
		\begin{split}
		(\mathcal{P}_2):\quad &\min\limits_{{\bf W}, {\bf a}, {\bf u}}\ \scaleobj{.8}{\sum_{\forall k}}\beta_k\left(u_ke_k-\log_2(u_k)\right)\\
		&{\rm s.t.}\ {\rm \textbf{Tr}}\left({\bf W}_s{\bf W}^H_s\right) \leq P_s,\ s\in\mathcal{S},\label{wmmse problem}
		\end{split}
	\end{align}
    where ${\bf a} \in \mathbb{C}^{K \times 1}$ and ${\bf u} \in \mathbb{C}^{K \times 1}$ are auxiliary variables, corresponding to the receive filters and weighting coefficients in the MSE formulation, respectively; while $e_k = \mathbb{E}_{{\bf H}, {\bf d}, n_k}\big[ | \hat{d}^{(m)}_k - d^{(m)}_k |^2 \big]$ denotes the MSE between the estimated symbol $\hat{d}^{(m)}_k$ and the transmit symbol $d^{(m)}_k$, which is given by
\begin{align}
    e_k &= a_ka^H_k\Big(\scaleobj{.8}{\sum_{i\neq k}}{\bf w}^H_{i}{\tilde {\boldsymbol{\Omega}}}_k{\bf w}_{i}+{\bf w}^H_{k}{\boldsymbol{\Omega}}_k{\bf w}_{k}\Big)- a^H_k\mathbb{E}\{{\bar {\bf h}}^T_{k}\}{\bf w}_{k} \notag\\
    &\quad - a_k{\bf w}^H_{k}\mathbb{E}\{{\bar {\bf h}}^*_{k}\}+ a_ka_k^H\sigma_k^2 + 1.
    \label{WMMSE simple}
\end{align}

    \begin{ppn}\label{no equivalence ppn}
    In the sense that the optimal ${\bf W}$ is identical, problems ($\mathcal{P}_1$) and ($\mathcal{P}_2$) are not equivalent when 
    \begin{align}
        \mathbb{E}\{{\bar {\bf h}}^*_{s,k}{\bar {\bf h}}^T_{s,k}\}\neq \mathbb{E}\{{\bar {\bf h}}^*_{s,k}\}\mathbb{E}\{{\bar {\bf h}}^T_{s,k}\},\ \forall s\in\mathcal{S}, k\in\mathcal{K}.
    \end{align}
    \end{ppn}
    \begin{pf}
    See Appendix \ref{no equivalence proof}.
    \end{pf}

    Then, the sCSI-related results in Section \ref{sum rate expression} yields
    \begin{align}
        \frac{(\mathbb{E}\{{\bar {\bf h}}^*_{s,k}{\bar {\bf h}}^T_{s,k}\})_{[i,j]}}{(\mathbb{E}\{{\bar {\bf h}}^*_{s,k}\}\mathbb{E}\{{\bar {\bf h}}^T_{s,k}\})_{[i,j]}} = \frac{\kappa_{s,k}+1}{\kappa_{s,k}} \frac{1}{|{\bar \varphi}_{s,k}|^2},\ \forall i,j.
    \end{align}
    Since $|{\bar \varphi}_{s,k}|^2 = \mathbb{E}\{|\varphi_{s,k}|^2\}-{\rm \textbf{var}}(\varphi_{s,k})=1-{\rm \textbf{var}}(\varphi_{s,k})$, we arrive at the following conclusion:
    \begin{remark}\label{equivalence remark}
        When the LoS path dominates the satellite channel and the phase-error variance is nearly zero, i.e., $\kappa_{s,k}\to\infty$ and ${\rm \textbf{var}}(\varphi_{s,k})\to 0$, ($\mathcal{P}_1$) and ($\mathcal{P}_2$) are nearly equivalent. Conversely, the greater the deviation between these two problems.
    \end{remark}

    \vspace{-3mm}
    
\subsection{Covariance Decomposition-based WMMSE Criterion}
    Thanks to~\eqref{WMMSE simple}, we introduce the following CDWMMSE formulation:
	\begin{align}
		\begin{split}
		(\mathcal{P}_3):\quad &\min\limits_{{\bf W}, {\bf A}, {\bf u}}\ \scaleobj{.8}{\sum_{\forall k}}\beta_k\left(u_k{\tilde e}_k-\log_2(u_k)\right)\\
		&{\rm s.t.}\ {\rm \textbf{Tr}}\left({\bf W}_s{\bf W}^H_s\right) \leq P_s,\ s\in\mathcal{S},\label{modified wmmse problem}
		\end{split}
	\end{align}
	where ${\tilde e}_k$ denotes the new MSE error, expressed as
	\begin{align}
		\begin{split}
        {\tilde e}_k 
        &\textstyle=  {\bf a}^H_k{\bf a}_k\Big(\scaleobj{.8}{\sum_{i\neq k}}{\bf w}^H_{i}{\tilde {\boldsymbol{\Omega}}}_k{\bf w}_{i}+{\bf w}^H_{k}{\boldsymbol{\Omega}}_k{\bf w}_{k}\Big)\\
        &\quad\   - {\bf a}^H_k{\bf Q}_{k}{\bf w}_{k} - {\bf w}^H_{k}{\bf Q}^H_{k}{\bf a}_k+ {\bf a}^H_k{\bf a}_k\sigma_k^2 + 1,
        \label{modified mse}
		\end{split}
	\end{align}
	where vector ${\bf Q}_{k}$ is constructed to satisfy ${\boldsymbol{\Omega}}_k = {\bf Q}^H_{k}{\bf Q}_{k}$, and ${\bf a}_k$ can be regarded as a virtual receiver. 
    \begin{ppn}\label{yes equivalence ppn}
   Problems ($\mathcal{P}_1$) and ($\mathcal{P}_3$) are equivalent in the sense that the optimal ${\bf W}$ is identical.
    \end{ppn}
    \vspace{-1mm}
    \begin{pf}
    See Appendix \ref{yes equivalence ppn proof}.
    \end{pf}
    \vspace{-3mm}
    \subsection{Low-Complexity Covariance Matrix Decomposition}
    \vspace{-1mm}
    The matrix ${\boldsymbol{\Omega}}_k$ is Hermitian and can be decomposed using Cholesky decomposition. However, this process incurs a computational cost on the order of $\mathcal{O}\left(S^3N_{\rm T}^3\right)$, and the matrix ${\bf Q}_k$ of size $SN_{\rm T} \times SN_{\rm T}$ further increases the complexity of subsequent optimization steps. Fortunately, as revealed by the analysis in Sections~\ref{channel model sec} and~\ref{sum rate expression}, multi-satellite channels exhibit distinctive statistical structures, which enable the following simplified decomposition:
    \begin{align}
        &{\boldsymbol{\Omega}}_k = {\bf Q}^H_{k}{\bf Q}_{k},\  {\bf Q}_{k} = 
        \begin{bmatrix}
            \mathbb{E}\{{\bar {\bf h}^T_k\}}\\
            {\bar {\bf Q}}_k\label{BMat eq1}
        \end{bmatrix}\in\mathbb{C}^{(S+1)\times SN_{\rm T}},\\
        &\ \ {\bar {\bf Q}}_k = {\rm \textbf{BlkDiag}}\{\varkappa_{1,k}{\bar {\bf h}}^T_{1,k}, ..., \varkappa_{S,k}{\bar {\bf h}}^T_{S,k}\},\label{BMat eq2}
    \end{align}
    where $\varkappa_{s,k} = \sqrt{\frac{\gamma_{s,k}}{\rho_{s,k}^*\rho_{s,k}}-1}$. 
    This decomposition does not incur additional computational burden, as the matrix ${\bf Q}_{k}$ has significantly fewer rows than columns, i.e., $S+1 \ll SN$, which facilitates the reduction of complexity in subsequent operations. In addition, to simplify the computation of the term ${\bf w}^H_{i}{\tilde {\boldsymbol{\Omega}}}_k{\bf w}_{i}$, we further factorize ${\tilde {\boldsymbol{\Omega}}}_k$ as ${\tilde {\boldsymbol{\Omega}}}_k = {\tilde {\bf Q}}^H_{k}{\tilde {\bf Q}}_k$, where ${\tilde {\bf Q}}_k \in \mathbb{C}^{S \times SN_{\rm T}}$ is defined by
    \begin{align}
        {\tilde {\bf Q}}_k = 
        {\rm \textbf{BlkDiag}}\{{\tilde \varkappa}_{1,k}{\bar {\bf h}}^T_{1,k}, ..., {\tilde \varkappa}_{S,k}{\bar {\bf h}}^T_{S,k}\},\label{tildeBMat eq1}
    \end{align}
    where ${\tilde \varkappa}_{s,k} = \sqrt{\frac{\gamma_{s,k}}{\rho_{s,k}^*\rho_{s,k}}}$. Under the above decomposition, the dimension of virtual receiver is ${\bf A}=[{\bf a}_1, ..., {\bf a}_K]\in\mathbb{C}^{(S+1)\times K}$.
    
\vspace{-3mm}
    \subsection{Multi-Satellite Distributed Precoding Design}\label{modified WMMME sec}
    \vspace{-1mm}
    Regarding per-satellite power constraints, our design is more complex than the single-satellite case, where solutions are more straightforward. Prior works \cite{10596023,10615897} simplify these constraints into per-user forms but neglect interference and lack rigorous constraint handling. To address $(\mathcal{P}_3)$, we adopt a similar transformation while ensuring a rigorous and accurate solution. Firstly, $(\mathcal{P}_3)$ is reformulated as:
    \vspace{-1mm}
    \begin{align}
        \begin{split}
        (\mathcal{P}_4):\quad &\min\limits_{{\bf W}, {\bf A}, {\bf u}}\ \scaleobj{.8}{\sum_{\forall k}}\beta_k\left(u_k{\tilde e}_k-\log_2(u_k)\right)\\
        &{\rm s.t.}\ {\rm \textbf{Tr}}\left({\bf w}_k{\bf w}^H_k\right) = P_k,\ k\in\mathcal{K}.\label{modified wmmse problem perUT}
        \end{split}
    \end{align}
    Here, the equality power constraint is imposed to simplify the optimization where the alternating methods, similar to those in \cite{christensen2008weighted, shi2011iteratively}, can be employed as detailed below.
    \subsubsection{Optimizing ${\bf A}$ and ${\bf u}$}
    When ${\bf W}$ and ${\bf u}$ are fixed, the optimal ${\bf A}$ can be obtained by taking the derivative of the objective function, given by:
    \vspace{-1mm}
    \begin{align}
        {\bf a}^{\star}_k=\frac{{\bf Q}_{k}{\bf w}_{k}}{{\bf w}^H_{k}{\boldsymbol{\Omega}}_k{\bf w}_{k}+\sum_{\forall i\neq k}{\bf w}^H_{i}{\tilde {\boldsymbol{\Omega}}}_k{\bf w}_{i} + \sigma^2_k}.
        \label{optimal receiver new}
    \end{align}
    Furthermore, when ${\bf W}$ and ${\bf A}$ are fixed, the optimal ${\bf u}$ is given by $u^{\star}_k={\tilde e}_k^{-1}$.
   \subsubsection{Optimizing ${\bf W}$}
  Given ${\bf A}$ and ${\bf u}$, solving $(\mathcal{P}_4)$ becomes slightly more challenging. Following \cite{christensen2008weighted}, we introduce a scaling factor $\eta_k$ for each ${\bf a}_k$
    to further optimize. This essentially corresponds to a joint optimization of ${\bf W}_k$ and part of ${\bf a}_k$. Then, the problem becomes
    \begin{align}
    \begin{split}
    &\min\limits_{{\bf W}, {\boldsymbol{\eta}}}\ f({\bf W}, {\boldsymbol{\eta}}) \qquad {\rm s.t.}\ {\rm \textbf{Tr}}\left({\bf w}_k{\bf w}^H_k\right) \leq P_k,\ k\in\mathcal{K}.
    \end{split}\label{problem for W new}
    \end{align}
    where the expression of $f({\bf W}, {\boldsymbol{\eta}})$ is \eqref{perUT W function1}. 
    Noting that SatCom systems are typically not interference-limited—i.e., interference is relatively weak compared to the desired signal and noise—we approximate $f({\bf W}, {\boldsymbol{\eta}})$ by ${\tilde f}({\bf W}, {\boldsymbol{\eta}})$, as defined in \eqref{perUT W function2}. This approximation decouples the optimization of ${\bf w}_k$ and $\eta_k$ for each UT, and the problem is reformulated as:
        \begin{figure*}[!t]
        \begin{align}
            f({\bf W}, {\boldsymbol{\eta}}) &\textstyle= \sum_{k=1}^{K}\beta_ku_k\left[ \frac{{\bf a}^H_k{\bf a}_k}{\eta^2_k}\left(\sum_{i\neq k}{\bf w}^H_{i}{\tilde {\boldsymbol{\Omega}}}_k{\bf w}_{i}+{\bf w}^H_{k}{\boldsymbol{\Omega}}_k{\bf w}_{k}\right)- \frac{{\bf a}^H_k{\bf Q}_{k}{\bf w}_{k}}{\eta_k} - \frac{{\bf w}^H_{k}{\bf Q}^H_{k}{\bf a}_k}{\eta_k}+ \frac{{\bf a}^H_k{\bf a}_k\sigma_k^2}{\eta^2_k}\right].
            \label{perUT W function1}\\
            {\tilde f}({\bf W}, {\boldsymbol{\eta}}) &\textstyle= \sum_{k=1}^{K}\beta_ku_k\left[ {\bf a}^H_k{\bf a}_k\left(\sum_{i\neq k}\frac{{\bf w}^H_{i}{\tilde {\boldsymbol{\Omega}}}_k{\bf w}_{i}}{\eta_i^2}+\frac{{\bf w}^H_{k}{\boldsymbol{\Omega}}_k{\bf w}_{k}}{\eta^2_k}\right)- \frac{{\bf a}^H_k{\bf Q}_{k}{\bf w}_{k}}{\eta_k} - \frac{{\bf w}^H_{k}{\bf Q}^H_{k}{\bf a}_k}{\eta_k}+ \frac{{\bf a}^H_k{\bf a}_k\sigma_k^2}{\eta^2_k}\right].
            \label{perUT W function2}
        \end{align}
        \vspace{-4mm} 
        \hrulefill  
    \end{figure*}
    \begin{align}
        &\min\limits_{{\bf w}_k, \eta_k}\ \frac{{\bf w}^H_k{\boldsymbol{\Xi}}_k{\bf w}_k}{\eta^2_k} -\frac{\beta_ku_k}{\eta_k}\left({\bf a}^H_k{\bf Q}_{k}{\bf w}_{k} \!+\! {\bf w}^H_{k}{\bf Q}^H_{k}{\bf a}_k \!-\! {\bf a}^H_k{\bf a}_k\sigma_k^2\right)\notag\\
        &\qquad{\rm s.t.}\ {\rm \textbf{Tr}}\left({\bf w}_k{\bf w}^H_k\right) = P_k,\label{perUT wmmse}
    \end{align}
    where ${\boldsymbol{\Xi}}_k = \beta_ku_k ({\bf a}^H_k{\bf a}_k){\boldsymbol{\Omega}}_k+\sum_{\forall i\neq k} \beta_iu_i ({\bf a}^H_i{\bf a}_i){\tilde {\boldsymbol{\Omega}}}_i$.

    \begin{ppn}\label{optimal perUT ppn}
    The following expressions of ${\bf w}_k$ and $\eta_k$ achieve the optimality of optimization problem \eqref{perUT wmmse}.
    \begin{align}
        &{\bf w}^{\star}_{k}
        = \eta^{\star}_k({\boldsymbol{\Xi}}_k+\frac{\beta_k u_k{\bf a}^H_k{\bf a}_k\sigma^2_k}{P_k}{\bf I})^{-1}\beta_k u_k{\bf Q}^H_{k}{\bf a}_k,\label{closed form new}\\
        &\textstyle\eta^{\star}_k = \sqrt{P_k/{\|({\boldsymbol{\Xi}}_k+\frac{\beta_k u_k{\bf a}^H_k{\bf a}_k\sigma^2_k}{P_k}{\bf I})^{-1}\beta_k u_k{\bf Q}^H_{k}{\bf a}_k\|^2_2}}.
    \end{align}
    \end{ppn}
    \begin{pf}
    See Appendix \ref{optimal perUT ppn proof}.
    \end{pf}

   By integrating results given in \eqref{optimal receiver new}, \eqref{modified mse}, and \ppnref{optimal perUT ppn}, the proposed algorithm, termed \textbf{MS-JoCDWM}, is summarized in \algref{MS-JoCDWM Algorithm}. The dominant computational cost arises from Step \ref{A1 closed form step}, with a complexity on the order of $\mathcal{O}\left(I_{\rm max}K(S^3N_{\rm T}^3)\right)$. Here, $I_{\rm max}$ denotes the maximum number of iterations, and $\chi$ represents the rate threshold used for convergence. Following the above method and Appendix \ref{no equivalence proof}, an analogous algorithm for the WMMSE problem (which is similar to that in \cite{10596023}) discussed in \secref{WMMME sec} can also be formulated, referred to as \textbf{MS-JoWM}.

    \begin{algorithm}[!t]
        \footnotesize
		\caption{MS-JoCDWM Algorithm for Problem \eqref{modified wmmse problem} }
		\label{MS-JoCDWM Algorithm}
		\begin{spacing}{1.2}
			\begin{algorithmic}[1]
				\STATE \textbf{Input:}$\{\!\gamma_{s,k}, \kappa_{s,k}, {\boldsymbol{\theta}}_{s,k}, {\bar \varphi}_{s,k}\!\}_{\forall s, k}$, $\{\!P_s\!\}_{\forall s}$, $\{\!\sigma^2_k, \beta_k\!\}_{\forall k}$, $I_{\rm max}$. \!\!\!\!\!\!
                \STATE Construct $\{{\bf Q}_k, {\tilde {\bf Q}}_k\}_{\forall k}$ using \eqref{BMat eq1}, \eqref{BMat eq2}, \eqref{barh eq1}, \eqref{tildeBMat eq1}.
                \STATE Initialize $\{{\bf W}_s\}_{\forall s}$. $n=0$.
                \STATE Initialize $P_k \!=\! (\sum_{s\in{\mathcal S}}P_s)/K$, ${\tilde e}_k \!=\! \chi$, ${\tilde e}'_k \!=\! \infty$, $\forall k\!\in\!{\mathcal K}$.
			\STATE \textbf{while} $n<I_{\rm max}$ \textbf{and} $\sum_{k=1}^K\beta_k\log_2({\tilde e}'_k/{\tilde e}_k)\!>\!\chi$ \textbf{do}
            \STATE \quad $n = n+1$.	${\tilde e}'_k = {\tilde e}_k$.
            \STATE \quad ${\bf a}_k=\frac{{\bf Q}_{k}{\bf w}_{k}}{\left(\sum_{i\neq k}{\bf w}^H_{i}{\tilde {\boldsymbol{\Omega}}}_k{\bf w}_{i}+{\bf w}^H_{k}{\boldsymbol{\Omega}}_k{\bf w}_{k}\right) + \sigma^2_k},\ \forall k\in\mathcal{K}$.
				\STATE \quad Compute ${\tilde e}_k$ with \eqref{modified mse}, $u_k={\tilde e}_k^{-1},\ \forall k\in\mathcal{K}$.
                \STATE \quad ${\boldsymbol{\Xi}}_k \!\!=\!\! \beta_ku_k ({\bf a}^H_k{\bf a}_k){\boldsymbol{\Omega}}_k+\!\!\!\sum\limits_{i=1,i\neq k}^{K} \beta_iu_i ({\bf a}^H_i{\bf a}_i){\tilde {\boldsymbol{\Omega}}}_i,\ \forall k\in\mathcal{K}$.
            \STATE \quad ${\bar {\bf w}}_{k}\!\!=\!\! \left({\boldsymbol{\Xi}}_k+\frac{\beta_k u_k{\bf a}^H_k{\bf a}_k\sigma^2_k}{P_k}{\bf I}\right)^{-1}\beta_k u_k{\bf Q}^H_{k}{\bf a}_k,\ \forall k\in\mathcal{K}$.
            \label{A1 closed form step}
                \STATE \quad $\eta_k = \sqrt{\frac{P_k}{\|{\bar {\bf w}}_{k}\|^2_2}},\ {\bf w}_{k}=\eta_k{\bar {\bf w}}_{k},\ \forall k\in\mathcal{K}$.
            \STATE  \textbf{end while}
            \STATE ${\bf W}'_{s} = [{\bf w}_{s,1},...,{\bf w}_{s,K}]$, 
                \STATE ${\bf W}_{s} = \min(\sqrt{\frac{P_s}{\|{\bf W}'_{s}\|^2_F}}, 1)\cdot {\bf W}'_{s}$, $\forall s\in \mathcal{S}$.
				\STATE \textbf{Output:}  $\{{\bf W}_s\}_{\forall s}$.
			\end{algorithmic}
		\end{spacing}
        \normalsize
	\end{algorithm}

    \vspace{-3mm}

    \section{Deep Learning-Aided Scalable Multi-Satellite Distributed Precoding Design}\label{modified WMMSE DL sec}

    Unlike terrestrial systems, SatComs require larger antenna arrays, i.e., a larger $N_{\rm T}$, due to their inherently power-limited nature to improve the link budget \cite{ASTSpaceMobile2023}. Furthermore, the number of users served by each satellite is significantly higher than that of terrestrial base stations, leading to a larger $K$. As a result, execution of \algref{MS-JoCDWM Algorithm} incurs high computational complexity. In this section, we propose a tiny neural network to directly obtain ${\bf A}$ and ${\bf u}$, thereby avoiding iterative computation of the algorithm. Based on our proof and exploitation of the key mapping properties in the precoding problem, this neural network can be trained in a specific scenario and applied in dynamic settings with varying numbers of satellites, users, and satellite antennas, achieving multidimensional scalability.

\vspace{-3mm}
    \subsection{Tensor Equivariance}\label{TE sec}
\vspace{-1mm}
    There are mainly two modeling approaches to leverage DL to exploit tensor equivariance (TE), where we use TE to refer to the extension encompassing higher-dimensional and higher-order equivariance and invariance \cite{wang2024towards, zaheer2017deep}. The first approach is Euclidean modeling, which directly derives the properties satisfied by the mapping and designs a matching network by focusing the processing along specific dimensions, as seen in \cite{wang2024towards, 9298921, 10584439}. The second approach is topological modeling, for example, constructing a bipartite graph topology and then designing a graph neural network \cite{9844981, bronstein2021geometric}. Although both approaches have their merits, we adopt the first method because of its intuitive design process.

    \subsubsection{Model Knowledge Discovery}
    We define ${\bar u}_k = u_k{\bf a}^H_k{\bf a}_k$ and ${\bar {\bf a}}_k = u_k{\bf a}_k$, thereby transforming \eqref{closed form new} into
    \begin{align}
        {\bf w}^{\star}_{k}
        = \eta^{\star}_k({\boldsymbol{\Xi}}_k+\frac{\beta_k {\bar u}_k\sigma^2_k}{P_k}{\bf I})^{-1}\beta_k {\bf Q}^H_{k}{\bar {\bf a}}_k.
        \label{closed form new2}
    \end{align}
    It can be observed that the core of \algref{MS-JoCDWM Algorithm} is to obtain the optimal ${\bar {\bf A}}\in\mathbb{C}^{K\times (S+1)}$ and ${\bar {\bf u}}\in\mathbb{R}^{K\times 1}$ through iterative computation, which is then used to calculate the precoding tensor $\tW\in\mathbb{C}^{K\times S\times N}$. The computation of ${\bar {\bf A}}$ and ${\bar {\bf u}}$ relies solely on $\{\!\gamma_{s,k}, \kappa_{s,k}, {\boldsymbol{\theta}}_{s,k}, {\bar \varphi}_{s,k}\!\}_{\forall s, k}$, $\{\!P_s\!\}_{\forall s}$, and $\{\!\sigma^2_k, \beta_k\!\}_{\forall k}$. 
    
    \subsubsection{Precoding Mapping Construction} For simplicity of expression, we define $\tD\in\mathbb{R}^{K\times S\times 5}$ and ${\bf F}\in\mathbb{R}^{K\times 2}$, and ${\bf p}=[P_1,...,P_S]^T\in\mathbb{R}^{S\times 1}$, where 
    \begin{align}
        \tD_{[k, s, :]} = [\gamma_{s,k}, \kappa_{s,k}, \theta^{\rm x}_{s,k}, \theta^{\rm y}_{s,k}, {\bar \varphi}_{s,k}],\ {\bf F}_{[k, :]} = [\sigma^2_k, \beta_k].
        \label{CSI for NN}
    \end{align}
    Then, we rearrange ${\bar {\bf A}}$ and ${\bar {\bf u}}$ as ${\tilde {\bf A}}\in\mathbb{C}^{K\times S}$ and ${\tilde {\bf U}}=[{\bar {\bf u}}, {\tilde {\bf a}}]\in\mathbb{C}^{K\times 2}$ where ${\tilde {\bf A}}$ and ${\tilde {\bf a}}$ are the two components that constitute ${\bar {\bf A}}$, i.e., ${\bar {\bf A}} = [{\tilde {\bf a}}, {\tilde {\bf A}}]$. Considering the non-convexity of problem \eqref{perUT wmmse}, the essence of \algref{MS-JoCDWM Algorithm} is a mapping from CSI $\tD$, ${\bf F}$, and ${\bf p}$ to an optimal pair of ${\tilde {\bf A}}$ and ${\tilde {\bf U}}$, i.e.,
    \begin{align}
        G(\tD, {\bf F}, {\bf p}) = {\tilde {\bf A}}^{\star},\ {\tilde {\bf U}}^{\star}.
    \end{align}

    \subsubsection{Proof of TE}
    We define $\langle\{{\tilde {\bf A}},{\tilde {\bf U}}\},\{\tD, {\bf F}, {\bf p}\}\rangle$ as a pairing of auxiliary variables and CSI for the closed-form expression 
    \eqref{closed form new2} to problem \eqref{modified wmmse problem}. The objective function achieved by $\tW = {\rm CFP}\left({\tilde {\bf A}},{\tilde {\bf U}}, \tD, {\bf F}, {\bf p}\right)$ and CSI $\{\tD, {\bf F}, {\bf p}\}$ in problem \eqref{modified wmmse problem} is denoted by $R\langle\{{\tilde {\bf A}},{\tilde {\bf U}}\},\{\tD, {\bf F}, {\bf p}\}\rangle$. We denote a permutation $\pi_M$ as a specific shuffling of the indices $[1,\dots,M]$ of a length-$M$ vector, and denote the set of all such permutations by $\mathbb{S}_M$. The operator $\circ_n$ indicates that the permutation acts on the $n$-th dimension of the tensor. For more details, see \cite{wang2024towards}.

    \begin{ppn}\label{ppn CF precoding}
    For any ${\pi_{S}}\in\mathbb{S}_{S}$ and ${\pi_{K}}\in\mathbb{S}_{K}$, we have
    \begin{align}
        \begin{split}
        &R\langle\{{\tilde {\bf A}},\ {\tilde {\bf U}}\},\{\tD, {\bf F}, {\bf p}\}\rangle\\
       &=R\langle\{\pi_{K}\!\circ_{1}\!{\tilde {\bf A}},\ \pi_{K}\!\circ_{1}\!{\tilde {\bf U}}\},\{\pi_{K}\!\circ_{1}\!\tD, \pi_{K}\!\circ_{1}\!{\bf F}, {\bf p}\}\rangle\\
        &=R\langle\{\pi_{S}\!\circ_{2}\!{\tilde {\bf A}},\ {\tilde {\bf U}}\},\{\pi_{S}\!\circ_{2}\!\tD, {\bf F}, \pi_{S}\circ_{1}{\bf p}\}\rangle.
        \end{split}
    \end{align}    
    \end{ppn}
    \begin{pf}
        For brevity, the proof is omitted since it is similar to that given in \cite{wang2024towards}.
    \end{pf}
    Based on the above proposition, it can be proven that the mapping has the following properties:
    \begin{align}
        G(\pi_{K}\!\circ_{1}\!\tD, \pi_{K}\!\circ_{1}\!{\bf F}, {\bf p}) &= \pi_{K}\!\circ_{1}\!{\tilde {\bf A}}^\star,\ \pi_{K}\!\circ_{1}\!{\tilde {\bf U}}^\star,\label{TE eq1}\\
        G(\pi_{S}\!\circ_{2}\!\tD, {\bf F}, \pi_{S}\circ_{1}{\bf p}) &= \pi_{S}\!\circ_{2}\!{\tilde {\bf A}}^\star,\ {\tilde {\bf U}}^\star.\label{TE eq2}
    \end{align}
    The above expression indicates that the mapping $G$ exhibits equivariance with respect to ${\tilde {\bf A}}$ and ${\tilde {\bf U}}$ along the user dimension, while it exhibits equivariance with respect to ${\tilde {\bf A}}$ and invariance with respect to ${\tilde {\bf U}}$ along the satellite dimension \cite{zaheer2017deep}. It is worth noting that, even though the problem is non-convex and multiple optimal pairs of ${\tilde {\bf A}}$ and ${\tilde {\bf U}}$ may exist, \eqref{TE eq1} and \eqref{TE eq2} still hold. A detailed analysis can be found in \cite{wang2024towards}.

\begin{figure*}[!t]
    \centering
    \includegraphics[width=6in]{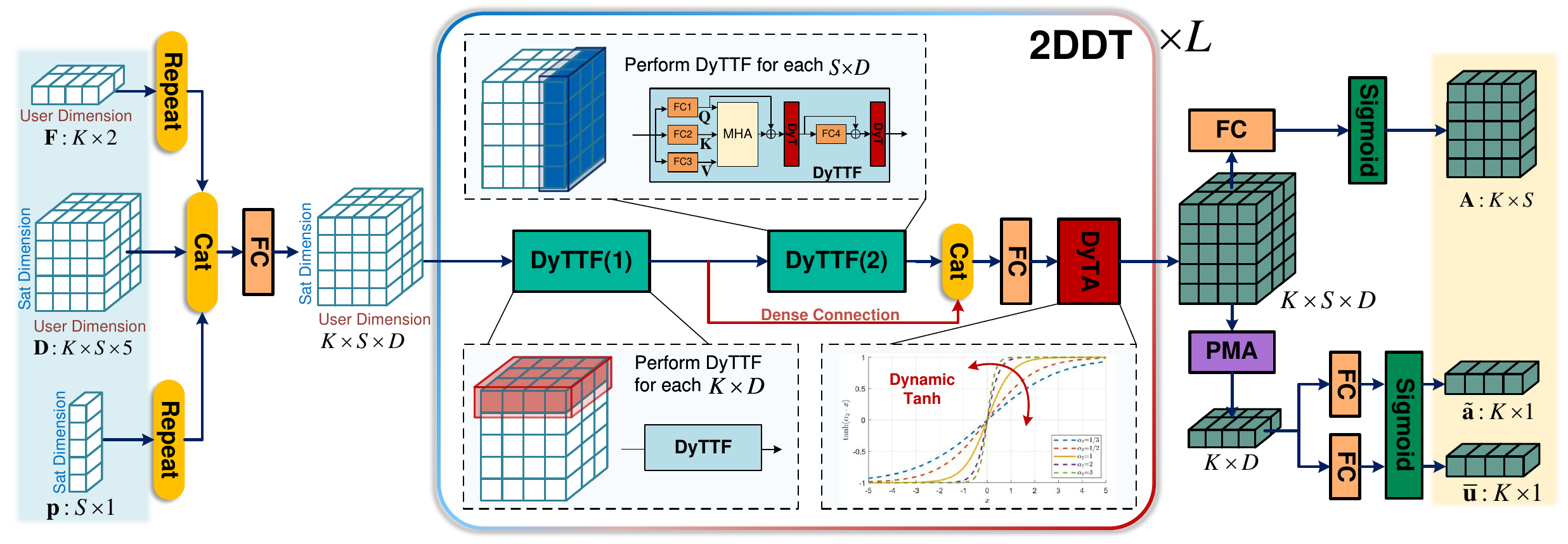}
    \vspace{-2mm}
    \caption{Overall architecture of the dense Transformer network.}
    \label{NN fig}
    \vspace{-3mm}
\end{figure*}

\vspace{-3mm}
\subsection{Dense Transformer Network}\label{NN design sec}
\vspace{-1mm}
    
    In this subsection, we design the network DTN to approximate the mapping $G$, ensuring that the network is structured to fully exploit the TE of $G$. We first propose the following $N$-dimensional dense Transformer $N{\rm D}{\rm DT}: \mathbb{R}^{M_1\times \cdots\times M_N\times D}\to \mathbb{R}^{M_1\times \cdots\times M_N\times D}$, which is applicable to mappings with TE in arbitrary dimensions. The operations performed by 
    $N{\rm D}{\rm DT}(\tX) = \tY$ are as follows.
    \begin{align}
        &\tX^{(n+1)} = {\rm DyTTF}^{(n)}\circ_{n} \tX^{(n)},\ \forall n\in\{1, ..., N-1\},\label{block equation 1}\\
        &\qquad\tY = {\rm DyTA}({\rm FC}([\tX^{(1)}, \tX^{(2)}, ..., \tX^{(N)}]_{N+1})),\label{block equation 2}
    \end{align}
    where $\tX^{(0)} = \tX$, and ${\rm FC}: \mathbb{R}^{ND}\to \mathbb{R}^{D}$ refers to a fully connected layer applied to the last dimension. The key components and mechanisms are given as follows:
    \begin{enumerate}
        \item \textbf{Transformer:} ${\rm DyTTF}: \mathbb{R}^{M\times D}\to \mathbb{R}^{M\times D}$represents a Transformer block \cite{vaswani2017attention, zhu2025transformersnormalization}. Due to the Transformer block's parameter-sharing mechanism, $M$ can adapt dynamically to the input size. Here, $\circ_n$ indicates the module operates on the $n$-th and the last dimensions, so the Transformer processes features of shape $M_n \times D$, treating other dimensions as batch size. The block uses an attention mechanism for efficient interactions among $M$ items while naturally preserving one-dimensional equivariance and delivering excellent performance \cite{yun2019transformers}. Its formulation can be founded in \cite{vaswani2017attention, zhu2025transformersnormalization}.
        \item \textbf{Dense Hierarchical Framework:} This framework is the core of the module. It sequentially applies different Transformers to each dimension as described in \eqref{block equation 1}, enabling feature interactions among dimensions while satisfying $N$-dimensional equivariance. Moreover, it aggregates the outputs of the DyT Transformers for different dimensions via dense connections \cite{huang2017densely}, as shown in  \eqref{block equation 2}, preserving key features from each stage to enhance performance.
        \item \textbf{Dynamic Tanh (DyT):} Unlike the layer normalization (LN) used in conventional Transformers, we replace it with a dynamic Tanh layer ${\rm DyT}$. This state-of-the-art technique reduces the training time and the computational cost of the normalization layer \cite{zhu2025transformersnormalization}. Specifically, the expression for ${\rm DyTA}: \mathbb{R}^{M_1\times \cdots\times M_N\times D}\to \mathbb{R}^{M_1\times \cdots\times M_N\times D}$ is given as follows:
        \begin{align}
            &{\rm DyTA}(\tX) = {\rm ACT}({\rm DyT}(\tX)),\\
            &{\rm DyT}(\tX)_{[m_1, ..., m_N, :]}={\rm DyT}(\tX_{[m_1, ..., m_N, :]}),\\
            &{\rm DyT}({\bf x})= {\boldsymbol{\alpha}}_1 \odot  {\rm tanh}(\alpha_2{\bf x})  +  {\boldsymbol{\alpha}}_3,
        \end{align}
        Where ${\rm ACT}$ represents an element-wise nonlinear activation function, ${\rm tanh}(x) = \frac{{\rm e}^x-{\rm e}^{-x}}{{\rm e}^x+{\rm e}^{-x}}$, and ${\boldsymbol{\alpha}}_1, {\boldsymbol{\alpha}}_3\in\mathbb{R}^{D}$, $\alpha_2\in\mathbb{R}$ are trainable parameters.
    \end{enumerate}
    It is worth noting that the original intent of this module design is not limited to the problem studied in this paper; it is also applicable for approximating other mappings that satisfy multidimensional equivariance \cite{wang2024towards, hartford2018deep}, thereby providing a new option for constructing low-complexity Transformer networks with equivariance in arbitrary dimensions.
    
    Based on the properties demonstrated in \secref{TE sec}, the mapping $G$ to be approximated exhibits corresponding TE in both the satellite and user dimensions. Therefore, we set $N=2$, i.e., we use $2{\rm D}{\rm DT}: \mathbb{R}^{M_1\times M_2\times D}\to \mathbb{R}^{M_1\times M_2\times D}$. Based on this module, following the guidance of \cite{wang2024towards}, we next complete the construction of the entire network.
    \subsubsection{Find TE} According to Section \ref{TE sec}, the mapping satisfys the equivariance in \eqref{TE eq1}-\eqref{TE eq2}.
    \subsubsection{Construct the Input}
	We construct the input as follows
\begin{align}
	\tX = [\tD,\tF,\tP]_{3}\in\mathbb{R}^{K\times S\times D_X},
\end{align}
where $D_X=8$. $\tF\in\mathbb{R}^{K\times S\times 2}$ and $\tP\in\mathbb{R}^{K\times S\times 1}$ are obtained by repeating ${\bf F}$ and ${\bf p}$, expressed as 
\begin{align}
    \tF_{[:, s, :]} = {\bf F},\ \tP_{[k, :, :]} = {\bf p},\ \forall s\in\mathcal{S}, \forall k\in\mathcal{K}.
\end{align}

\subsubsection{Build Equivariant Network}
In addition to processing the inputs and outputs, the network primarily comprises $L$ $2{\rm D}{\rm DT}$ modules that facilitate feature interactions across all dimensions of the processed input tensor $\tX$, i.e., 
\begin{align}
    \tM = 2{\rm D}{\rm DT}^{\times L}({\rm FC}(\tX))\in\mathbb{R}^{K\times S\times D},
\end{align}
    where we use ${\times L}$ to denote the stacking of $L$ blocks.
\begin{flalign}
 &\ \tY_1 \!=\! {\rm FC}_{{\bar {\bf A}}}(\tM)\!\in\!\mathbb{R}^{K\!\times\! S \!\times\! 2},\ {\bf Y}_2 \!=\! {\rm PMA}\!\circ_{2}\!(\tM)\!\in\!\mathbb{R}^{K\!\times\! D},&
 \end{flalign}
where ${\rm PMA}: \mathbb{R}^{M\times D}\to \mathbb{R}^{D}$ is a one-dimensional invariant module constructed based on the Transformer. It features the parameter-sharing mechanism, allowing $M$ to vary dynamically. Its specific expression can be found in \cite{wang2024towards,10584439,lee2019set}. 

\begin{remark}
It is easy to see that the processing from $\mathbf{D}$, $\mathbf{F}$, and $\mathbf{p}$ to $\mathbf{Y}_1$ and $\mathbf{Y}_2$ satisfies the TE \eqref{TE eq1} and \eqref{TE eq2}. 
\end{remark}

\subsubsection{Design The Output Layer} The final results can be 
 \begin{align}
&{\tilde {\bf A}} = \tY_{1[:, 
:, 1]} + j\tY_{1[:, :, 2]},\ {\bar {\bf u}} = {\rm SIG}({\rm FC}_{{\bar {\bf u}}}({\bf Y}_2)),\\
&{\tilde {\bf a}} = {\rm SIG}({\rm FC}_{{\tilde {\bf a}}}({\bf Y}_2))_{[:, 1]} + j{\rm SIG}({\rm FC}_{{\tilde {\bf a}}}({\bf Y}_2))_{[:, 2]},
\end{align}
where ${\rm SIG}(x) = \frac{1}{1+{\rm e}^{-x}}$ is the sigmoid function.

In summary, the overall architecture of DTN is shown in \figref{NN fig}. Based on the network architecture, our proposed DL algorithm, MS-JoCDWM-DL, is presented as \algref{MS-JoCDWM-DL Algorithm}. The computational complexity of the algorithm mainly lies in steps \ref{A2 net step} and \ref{A2 closed-form step}, resulting in a total complexity of $\mathcal{O}\left(K(S^3N_{\rm T}^3)+LD(K^2+S^2)\right)$.  In fact, since $N_{\rm T}\gg K\gg S$, the second part of the complexity can be neglected. Since the designed network exploits the mapping properties discussed in \secref{TE sec} and incorporates a parameter-sharing mechanism \cite{wang2024towards}, the total number of parameters is $\mathcal{O}\left(D^2\right)$, independent of $S$, $K$, and $N_{\rm T}$. Furthermore, the parameter-sharing mechanism enables the network trained under specific values of $S$ and $K$ to be applied to scenarios with different $S$ and $K$. Moreover, since the network only processes sCSI and does not involve the dimension $N_T$, it can also be applied to scenarios with different $N_{\rm T}$. This generalizability with respect to $S$, $K$, and $N_{\rm T}$ is referred to as `3D scalability'.

        \begin{algorithm}[t]
        \footnotesize
		\caption{MS-JoCDWM-DL Algorithm}
		\label{MS-JoCDWM-DL Algorithm}
		\begin{spacing}{1.2}
			\begin{algorithmic}[1]
				\STATE \textbf{Input:}$\{\!\gamma_{s,k}, \kappa_{s,k}, {\boldsymbol{\theta}}_{s,k}, {\bar \varphi}_{s,k}\!\}_{\forall s, k}$, $\{\!P_s\!\}_{\forall s}$, $\{\!\sigma^2_k, \beta_k\!\}_{\forall k}$. \!\!\!\!\!\!
                \STATE Construct $\{{\bf Q}_k, {\tilde {\bf Q}}_k\}_{\forall k}$ using \eqref{BMat eq1}, \eqref{BMat eq2}, \eqref{barh eq1}, \eqref{tildeBMat eq1}.
                \STATE Construct $\tD, {\bf F}, {\bf p}$ using \eqref{CSI for NN}.
                \STATE $P_k \!=\! (\sum_{s\in{\mathcal S}}P_s)/K$, $\forall k\!\in\!{\mathcal K}$.
                \STATE ${\tilde {\bf A}}, {\tilde {\bf a}}, {\bar {\bf u}} = {\rm DTN}(\tD, {\bf F}, {\bf p})$,\  ${\bar {\bf A}} = [{\tilde {\bf a}}, {\tilde {\bf A}}]$. \label{A2 net step}
			\STATE ${\bar {\bf w}}_{k}
        = ({\boldsymbol{\Xi}}_k+\frac{\beta_k {\bar u}_k\sigma^2_k}{P_k}{\bf I})^{-1}\beta_k {\bf Q}^H_{k}{\bar {\bf a}}_k,\ \forall k\in\mathcal{K}$. \label{A2 closed-form step}
                \STATE $\eta_k = \sqrt{\frac{P_k}{\|{\bar {\bf w}}_{k}\|^2_2}},\ {\bf w}_{k}=\eta_k{\bar {\bf w}}_{k},\ \forall k\in\mathcal{K}$.
            \STATE ${\bf W}'_{s} = [{\bf w}_{s,1},...,{\bf w}_{s,K}]$, 
                \STATE ${\bf W}_{s} = \min(\sqrt{\frac{P_s}{\|{\bf W}'_{s}\|^2_F}}, 1)\cdot {\bf W}'_{s}$, $\forall s\in \mathcal{S}$.
				\STATE \textbf{Output:}  $\{{\bf W}_s\}_{\forall s}$.
			\end{algorithmic}
		\end{spacing}
        \normalsize
	\end{algorithm}
\vspace{-3mm}
    \subsection{Scaling-Invariant Loss for Supervised Learning}
    In previous model-driven DL methods, unsupervised learning was typically used \cite{wang2024towards}, requiring the GPU to batch-compute closed-form precoding \eqref{closed form new2} and gradients for backpropagation. However, due to the significantly larger matrix dimensions in SatComs, this approach demands enormous memory and computational resources—there is a clear need for supervised learning, which nonetheless brings its own challenge: choosing an appropriate loss function. Next, we analyze the limitations of commonly used loss functions MSE and NMSE; then propose a loss function specifically tailored to the considered problem.

    We rearrange matrices $\{{\tilde {\bf A}},\ {\tilde {\bf U}}\}$ into $\{{\bar {\bf A}}, {\bar {\bf u}}\}$, and redefine the objective function achieved by $\tW = {\rm CFP}\left({\bar {\bf A}}, {\bar {\bf u}}, \tD, {\bf F}, {\bf p}\right)$ and CSI $\{\tD, {\bf F}, {\bf p}\}$ in problem \eqref{modified wmmse problem} is denoted by $R\langle\{{\bar {\bf A}}, {\bar {\bf u}}\},\{\tD, {\bf F}, {\bf p}\}\rangle$. 
    \begin{remark}
        It is easy to verify
        \vspace{-1mm}
        \begin{align}
            R\langle\{{\bar {\bf A}}, {\bar {\bf u}}\},\{\tD, {\bf F}, {\bf p}\}\rangle\!=R\langle\{{\boldsymbol{\Gamma}}_{a}{\bar {\bf A}}, \gamma_{u}\!\cdot\!{\bar {\bf u}}\},\{\tD, {\bf F}, {\bf p}\}\rangle.
            \vspace{-1mm}
        \end{align}
        where ${\boldsymbol{\Gamma}}_{a} = {\rm diag}\{\gamma_{a, 1}, ..., \gamma_{a, K}\}$. $\gamma_{a, 1}, ..., \gamma_{a, K}, \gamma_{u}\in\mathbb{C}$ represent nonzero scaling factors. This implies that scaling ${\bar {\bf a}}_k$ and ${\bar {\bf u}}$ separately does not affect the sum rate.
    \end{remark}
    Clearly, neither the MSE nor the NMSE loss is suitable for measuring the distance between the network output and the label of ${\bar {\bf A}}$ and ${\bar {\bf u}}$, as both are significantly affected by scaling. To this end, we define the following scale-invariant NMSE function as the loss function for supervised learning.
    \vspace{-1mm}
    \begin{align}
        &\textstyle\ell_{\rm SIN}({\hat {\bf X}}, {\bf X}) = \frac{\|{\bf X}-\xi^\star{\hat {\bf X}}\|_{\rm F}^2}{\|{\bf X}\|_{\rm F}^2+\epsilon},\\
        &\textstyle\xi^\star = \arg\min_{\xi}\|{\bf X}-\xi{\hat {\bf X}}\|_{\rm F} = \frac{\langle {\hat {\bf X}}, {\bf X} \rangle_F}{\langle {\bf X}, {\bf X} \rangle_F}\in\mathbb{C},
        \vspace{-1mm}
    \end{align}
    where $\epsilon=1 \times 10^{-8}$ is a constant set to prevent the denominator from becoming too small. 
    Another option is $\textstyle{\text{VCOSSIM}}({\hat {\bf X}}, {\bf X}) = \frac{|{\rm vec}({\hat {\bf X}})^H{\rm vec}({\bf X})|}{\|{\rm vec}({\hat {\bf X}})\|\|{\rm vec}({\bf X})\|}$, which computes cosine similarity after vectorizing the matrix.
    Taking $\text{SI-NMSE}$ as an example, the loss function for training the network in \secref{NN design sec} is defined as follows:
    \vspace{-1mm}
    \begin{align}
        &{\rm Loss} = -\frac{1}{N_{\rm sp}}\scaleobj{.8}{\sum_{n=1}^{N_{\rm sp}}}c \Big(\frac{1}{K}\scaleobj{.8}{\sum_{k=1}^K} \ell_{\rm SIN}({\bar {\bf a}}_k[n], {\bar {\bf a}}_k^\star[n])\Big)\notag\\
        &\qquad\qquad\qquad\qquad\qquad+(1-c) \ell_{\rm SIN}({\bar {\bf u}}[n], {\bar {\bf u}}^\star[n]).
        \vspace{-1mm}
    \end{align}
    where the index $n$ refers to the $n$-th sample, and $c \in (0,1)$ is a constant used to adjust the weighting of the losses, which is set to 0.5 in simulations.

    \vspace{-3mm}
    \subsection{Dataset and Training Details}
    The dataset comprises $5000$ samples, with $4500$ used for training, $500$ for validation, and an additional $50$ for testing. Each sample randomly selects the center of the cooperative transmission region within the global coverage of the constellation, generates users, and determines the cooperating satellites, consistent with the Monte Carlo settings described in \secref{Section 6}.
    It further includes data corresponding to seven transmit power levels, namely $P_{\rm s}\in[20, 25, 30, 35, 40, 45, 50]{\rm dBm}$, which are randomly selected during training to enable the network to operate under various link budgets. The number of iterations and the batch size are set to $15000\sim20000$ and $500$, respectively. The Adam optimizer is used with a learning rate of \(5 \times 10^{-4}\). The versions of PyTorch and torchvision are 1.12.1 and 0.13.1, respectively.

    \vspace{-3mm}
    \section{Numerical Results}\label{Section 6}
\begin{figure}[t]
    \centering
    \begin{minipage}{1\linewidth}
        \centering
        \includegraphics[width=0.8\linewidth]{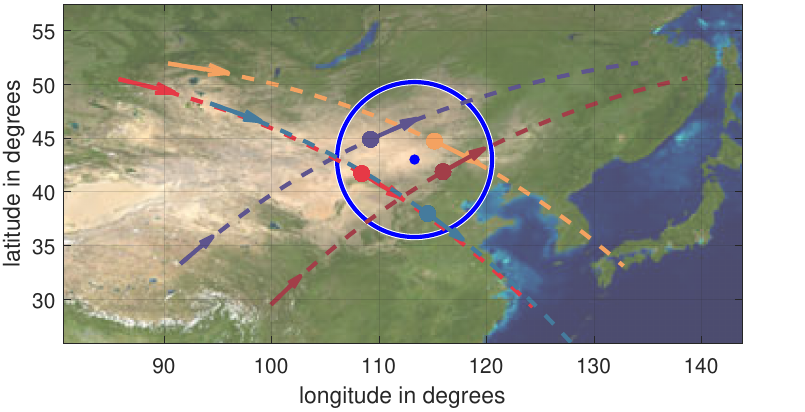}
        \vspace{-2mm}
        \caption{2D visualization of satellites in one of the sampled scenarios.}
        \label{earth satellite 2D}
    \end{minipage}
\end{figure}

    We use the QuaDRiGa channel simulator to generate the scenario and radio channel parameters \cite{6902008,9815679,QuaDRiGa2023}. In particular, the channel parameters are generated with the simulator under its `QuaDRiGa\_NTN\_Urban\_LOS' scenario \cite{6902008}. This simulator, with appropriately calibrated parameters, is aligned with the channel model considered in this work and the Third Generation Partnership Project specifications \cite{3GPP_TR_38_811, 9815679}. Other simulation parameters are provided in Table \ref{Simulation Parameter Table}. Although the radius of the cooperative transmission coverage area is set at 800 km, it does not affect the performance conclusions of the designed method. Monte Carlo simulations are conducted, where each run involves randomly selecting a point within the constellation's coverage area as the center. A circular region with the given coverage radius is defined, and the $S$ closest satellites to the center are selected for cooperative transmission. Under a specific random seed, the selected service area and satellites' 2D visualizations are shown in \figref{earth satellite 2D}. In our simulations, we assume that the power constraints of all satellites are identical, i.e., $P_s = P_{\rm TX}, \forall s$, and set its range as $P_{\rm TX}\in[20{\rm dBm}, 50{\rm dBm}]$. This range can also simulate different link budgets resulting from various transceiver configurations \cite{3gpp_tr_38_821}. Besides, we use $R_{\rm E} = \sum_{k\in\mathcal{K}}\mathbb{E}\{R_k\}$ to evaluate the performance. As stated in \secref{Interference sec}, the distribution of the phase error $\varphi_{s,k}$ depends on the estimation method, which is beyond the scope of this paper. Without loss of generality, we adopt the modeling approach used in \cite{10596023, wang2021resource}, where $\varphi_{s,k} = {\rm e}^{j\varrho_{s,k}}$ and $\varrho_{s,k} \sim \mathcal{N}(0, \varsigma^2_{s,k})$.

   \begin{table}[!t]
    \centering
    \caption{Simulation Parameters \cite{3gpp_tr_38_821, 3GPP_TR_38_811, SpaceX_Gen2_2021, 9998075, 9815679, 10440321}}
    \vspace{-2mm}
     \resizebox{0.35\textwidth}{!}{%
    \begin{tabular}{ll}  
    \toprule
    \textbf{Parameter} & \textbf{Value} \\ 
    \midrule
    Satellites altitude & 600 km \\ 
    Carrier frequency & 2 GHz \\ 
    System Bandwidth (DL) & 20 MHz \\ 
    Subcarrier Spacing & 30kHz\\
    UT Noise figure & 7 dB \\ 
    UT Antenna temperature & 290 K \\ 
    Coverage radius & 800 km \\ 
    Number of UTs & 12 - 48 \\ 
    Number of cooperating satellites & 5\\
    Distribution of UTs & Uniform \\
    \midrule
    Per-element gain of TX antennas & 6dBi \\ 
    Gain of RX antennas & 0dBi \\
    Transmit antenna size & $N_{\rm v}=N_{\rm h}=10$\\
    \midrule
    Constellation Type & Walker-Delta \\ 
    Orbital Planes & 28 \\ 
    Satellites Per Plane & 60 \\ 
    Inclination (degrees) & 53 \\ 
    \bottomrule
    \end{tabular}}
    \label{Simulation Parameter Table}
    \vspace{-3mm}
\end{table}

This section compares the following schemes:
\begin{itemize}
\item `\textbf{SS-M}' and `\textbf{SS-WM}': The MMSE and WMMSE precoding performed  based on the expectation of the compensated channel, with service provided by single satellite \cite{christensen2008weighted}. 
\item `\textbf{MS-SepWM}': WMMSE precoding is performed separately on multiple satellites based on the expectation of the compensated channels.
\item `\textbf{MS-JoWM}' and `\textbf{MS-JoCDWM}': The proposed WMMSE- and CDWMMSE-based method in \secref{WMMME sec} and \ref{modified WMMME sec} with $I_{\rm max}=10$ and $\chi=0.001$.
\item `\textbf{MS-JoCDWM-DL}': The proposed DL-based method for CDWMMSE in \secref{modified WMMSE DL sec} with $L=3$ and $D=32$.
\end{itemize}

\begin{figure}[t]
    \centering
    \includegraphics[width=2.7in]{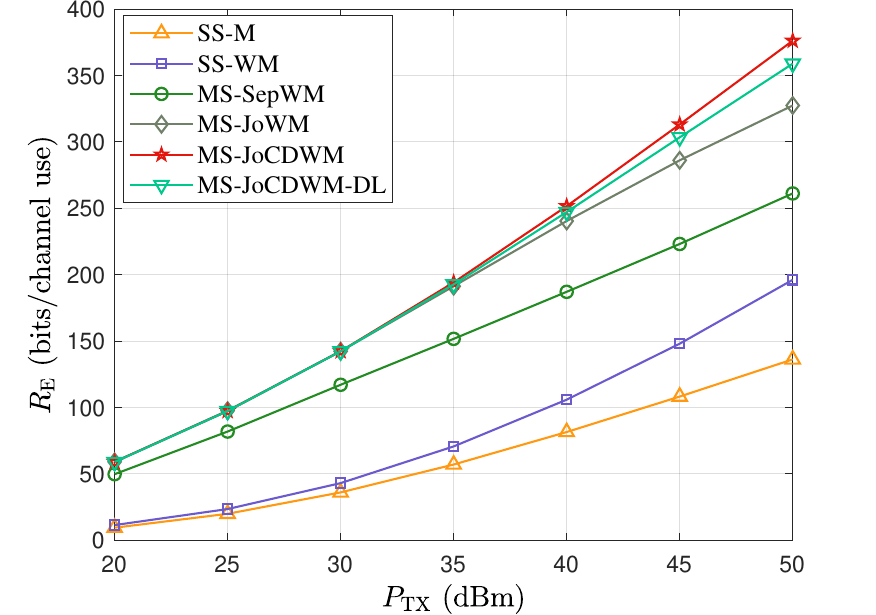}
    \vspace{-2mm}
    \caption{Average sum rate vs $P_{\rm TX}$, $\zeta^2_{s,k} = 0.05$.}
    \label{sumrate_snr_zetap05 fig}
    \vspace{-3mm}
\end{figure}

\begin{figure}[t]
    \centering
    \includegraphics[width=2.7in]{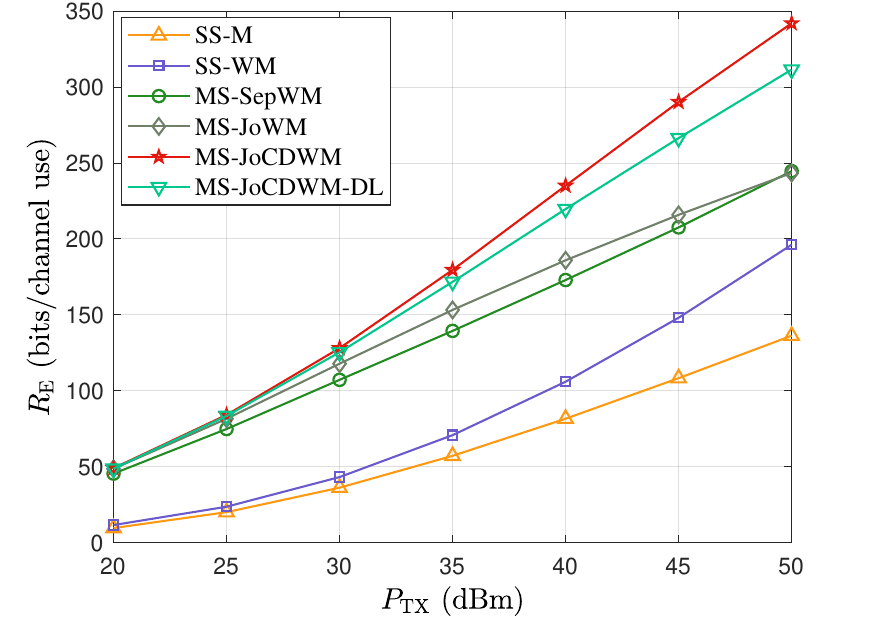}
    \vspace{-2mm}
    \caption{Average sum rate vs $P_{\rm TX}$, $\zeta^2_{s,k} = 0.5$.}
    \label{sumrate_snr_zetap5 fig}
    \vspace{-3mm}
\end{figure}

\figref{sumrate_snr_zetap05 fig} and \figref{sumrate_snr_zetap5 fig} compare the sum rate performance of each method under varying transmit power for the cases of $\zeta^2_{s,k} = 0.05$ and $\zeta^2_{s,k} = 0.5$ with $K=48$, respectively. It can be observed that, overall, multi-satellite transmission outperforms single-satellite transmission. Among the multi-satellite approaches, MS-JoCDWM outperforms MS-JoCDWM-DL, which in turn outperforms MS-JoWM. Moreover, as $\zeta^2_{s,k}$ increases, the performance gap between MS-JoCDWM and MS-JoWM becomes more pronounced. Although MS-JoCDWM-DL performs slightly worse than MS-JoCDWM, it achieves even much lower computational complexity than MS-JoWM due to its non-iterative nature, highlighting the advantage of DL-based approach.


\begin{figure}[t]
    \centering
    \includegraphics[width=2.7in]{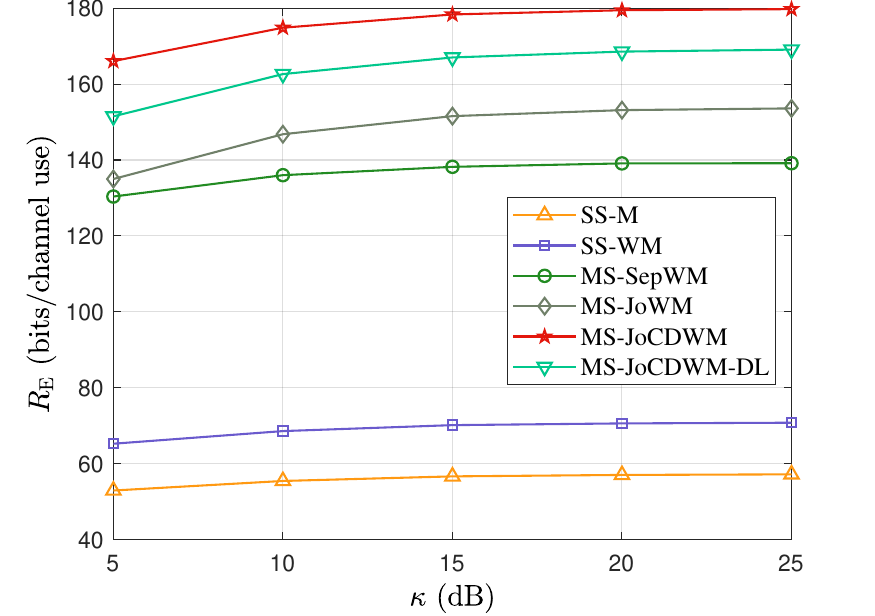}
    \vspace{-2mm}
    \caption{Average sum rate vs $\kappa_{s,k}$, $\zeta^2_{s,k} = 0.5$, $P_{\rm TX}=35{\rm dBm}$.}
    \label{sumrate_kappa_35dBm_zetap5 fig}
    \vspace{-3mm}
\end{figure}


\begin{figure}[t]
    \centering
    \includegraphics[width=2.7in]{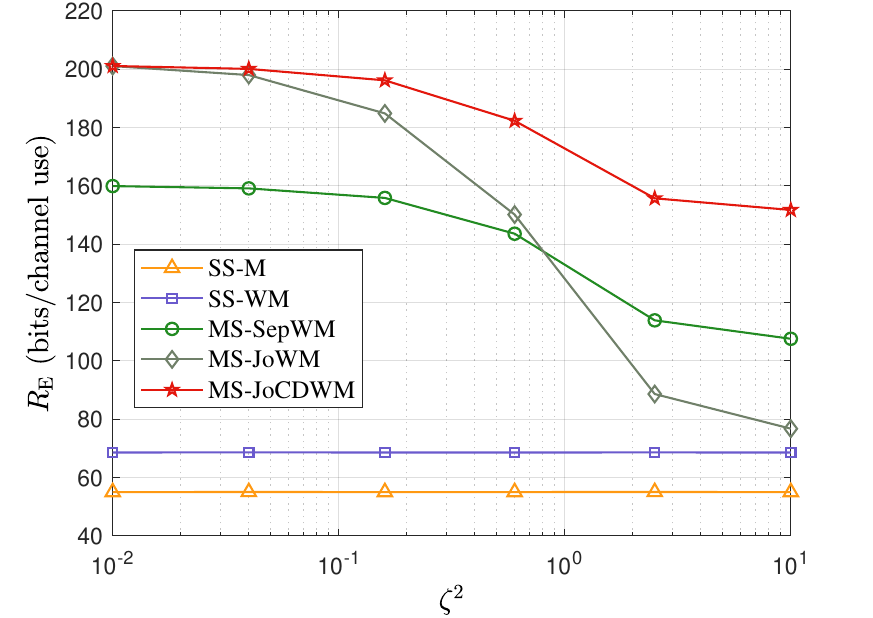}
    \vspace{-2mm}
    \caption{Average sum rate vs $\zeta^2_{s,k}$, $\kappa_{s,k} = 25{\rm dB}$, $P_{\rm TX}=35{\rm dBm}$.}
    \label{sumrate_zetap_35dBm_kappa25 fig}
    \vspace{-3mm}
\end{figure}

\figref{sumrate_kappa_35dBm_zetap5 fig} and \figref{sumrate_zetap_35dBm_kappa25 fig} evaluate the impact of parameters $\kappa_{s,k}$ and $\zeta^2_{s,k}$ on the performance of the proposed methods, respectively. Although $\kappa_{s,k}$ in the channel parameters is originally generated by QuaDRiGa, we manually control it here to enable a clearer comparison. As shown in \figref{sumrate_kappa_35dBm_zetap5 fig}, the performance of all methods degrades as $\kappa$ decreases. This is because a smaller $\kappa$ implies a stronger NLoS component in the channel, leading to higher channel variance, which degrades the effectiveness of precoding based on statistical information. In \figref{sumrate_zetap_35dBm_kappa25 fig}, the performance of each method gradually decreases with increasing $\zeta^2_{s,k}$, since a larger $\zeta^2_{s,k}$ results in lower signal coherence at the receiver. Nevertheless, the multi-satellite schemes still outperform the single-satellite scheme, as distributed precoding not only enables coherent transmission but also enhances spatial multiplexing capability at the transmitter.

    \begin{figure}[t]
        \centering
        \includegraphics[width=2.7in]{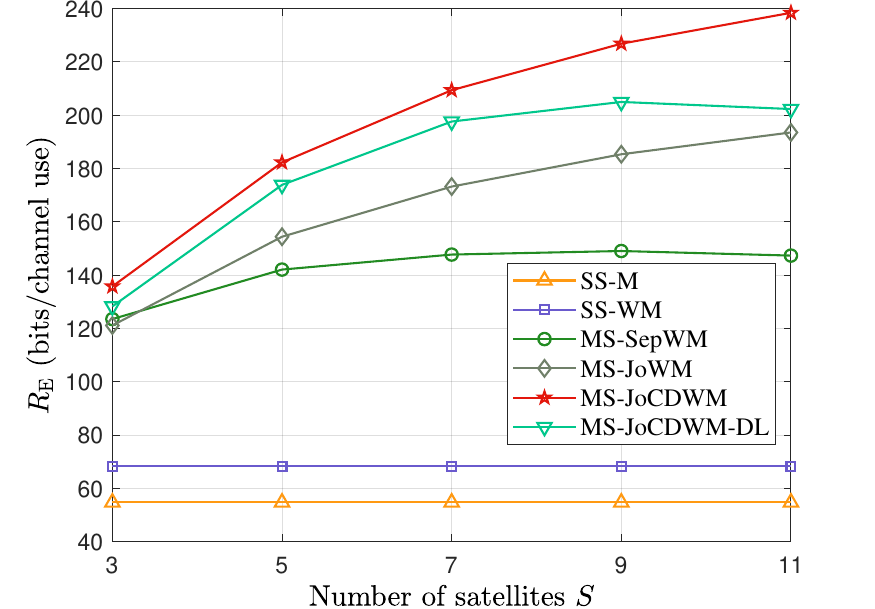}
        \vspace{-2mm}
        \caption{Average sum rate vs $S$, $\zeta^2_{s,k} = 0.5$, $P_{\rm TX}=35{\rm dBm}$, with MS-JoCDWM-DL trained under $S=5$.}
        \label{sumrate_allSats_zetap5_35dBm fig}
        \vspace{-3mm}
    \end{figure}

    \begin{figure}[t]
        \centering
        \includegraphics[width=2.7in]{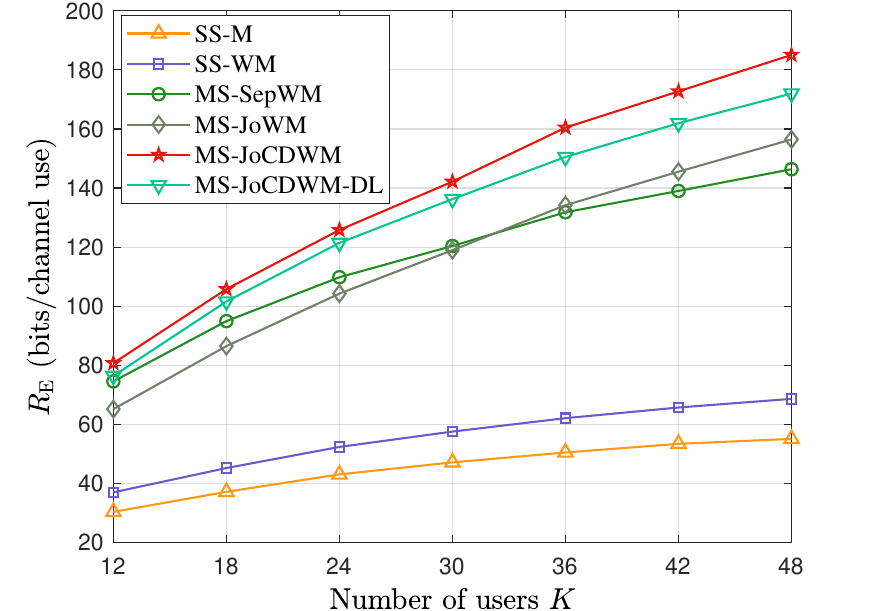}
        \vspace{-2mm}
        \caption{Average sum rate vs $K$, $\zeta^2_{s,k} = 0.5$, $P_{\rm TX}=35{\rm dBm}$, with MS-JoCDWM-DL trained under $K=48$.}
        \label{sumrate_kNum_zetap5 fig}
    \end{figure}

\figref{sumrate_allSats_zetap5_35dBm fig} and \figref{sumrate_kNum_zetap5 fig} show the impact of the number of cooperative satellites and the number of users on performance, respectively. As shown in \figref{sumrate_allSats_zetap5_35dBm fig}, the performance gain of the proposed methods over single-satellite transmission increases progressively with the number of cooperative satellites. Notably, since MS-SepWM does not suppress interference cooperatively, increasing the number of satellites may introduce more interference, ultimately leading to performance degradation. As shown in \figref{sumrate_kNum_zetap5 fig}, the proposed MS-JoCDWM consistently achieves the best performance across different numbers of users. In contrast, MS-JoWM, due to limitations in its design criterion, only demonstrates performance advantages when the number of users becomes large. Meanwhile, \figref{sumrate_allSats_zetap5_35dBm fig} and \figref{sumrate_kNum_zetap5 fig} demonstrate the excellent generalization capability of the proposed method MS-JoCDWM-DL. The network trained under the scenario with $S = 5$ and $K=48$ can be deployed across a wide range of $S$ and $K$ values while still maintaining outstanding performance.

\begin{table}[t]
 \centering
 \caption{Computational Complexity of Precoding Schemes}
 \vspace{-2mm}
 \resizebox{0.9\columnwidth}{!}{
  \begin{tabular}{ccc}
  \toprule
  Methods & Complexity Order \\
  \midrule
  SS-M &  ${\mathcal O}(N_{\rm T}K^2)$   \\
  SS-WM &  ${\mathcal O}(I_{\rm max1}N_{\rm T}K^2)$   \\
  MS-SepWM &  ${\mathcal O}(I_{\rm max1}N_{\rm T}K^2S)$   \\
  MS-JoWM &  ${\mathcal O}(I_{\rm max}N_{\rm T}^3KS^3+I_{\rm max1}N_{\rm T}K^2S)$   \\
  MS-JoCDWM &  ${\mathcal O}(I_{\rm max}N_{\rm T}^3KS^3+I_{\rm max1}N_{\rm T}K^2S)$   \\
  MS-JoCDWM-DL &  $\mathcal{O}\left(N_{\rm T}^3 K S^3 + LD(K^2+S^2)\right)$  \\
  \bottomrule
  \end{tabular}
  }
 \label{Complexity Table}%
 \vspace{-3mm}
\end{table}

\begin{figure}[t]
    \centering
    \includegraphics[width=2.7in]{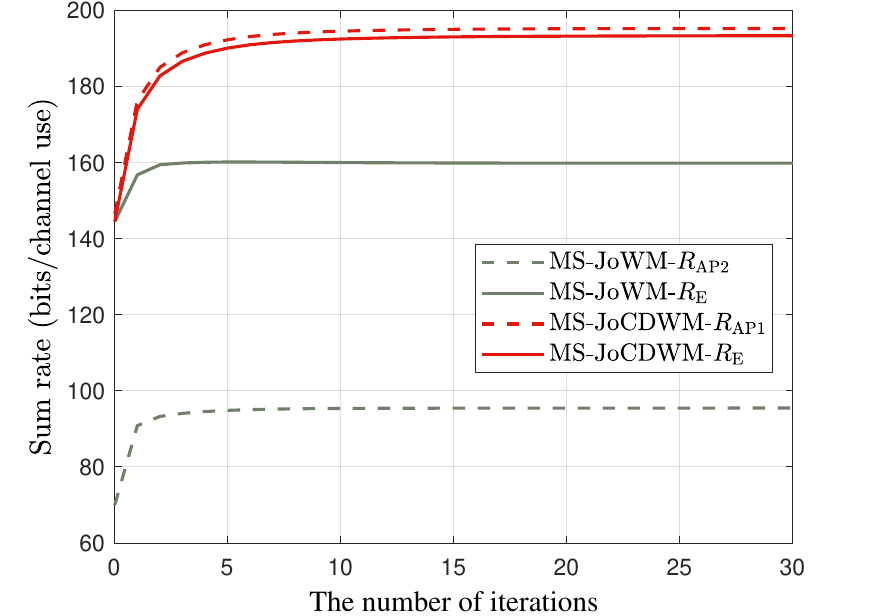}
    \vspace{-2mm}
    \caption{Convergence of the proposed method, $\zeta^2_{s,k} = 0.5$, $P_{
    \rm TX}=35{\rm dBm}$.}
    \label{iter fig}
    \vspace{-3mm}
\end{figure}

 \figref{iter fig} shows the evolution of \(R_{\rm E}\) along with two sum-rate approximations over the iterations: $R_{\rm AP1} = \sum_{k\in\mathcal{K}} \bar{R}_k$
and $R_{\rm AP2} = \sum_{k\in\mathcal{K}} \tilde{R}_k$, where $\tilde{R}_k = \log_2\left(e^{-1}_k\right)$. These are sum rate approximations adopted by MS-JoCDWM and MS-JoWM, respectively. It is evident that $R_{\rm AP2}$ deviates further from $R_{\rm E}$, while $R_{\rm AP1}$ is closer to $R_{\rm E}$—this is one of the reasons why \ppnref{no equivalence ppn} and \textbf{\ref{yes equivalence ppn}} demonstrate the superiority of MS-JoCDWM over MS-JoWM.

\begin{figure}[t]
    \centering
    \includegraphics[width=2.7in]{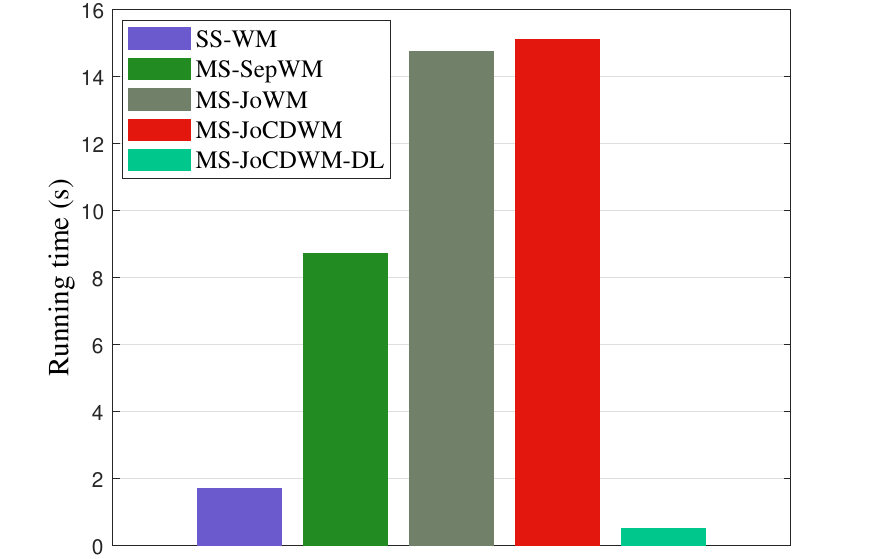}
    \vspace{-2mm}
    \caption{Running time comparison.}
    \label{time fig}
    \vspace{-3mm}
\end{figure}

Table \ref{Complexity Table} compares the computational complexity of the considered algorithms. 
The complexity of the baselines have been reduced using the matrix inversion lemma and therefore does not involve $N^3_{\rm T}$, and $I_{\rm max1}=300$ is the maximum number of iterations. It can be observed that algorithms involving multi-satellite cooperation incur higher computational complexity than single-satellite algorithms due to the need for joint optimization, which is consistent with the results in \cite{10596023}. Among the cooperative algorithms, although MS-JoCDWM achieves significantly better performance than MS-JoWM—as will be shown in the subsequent results—it exhibits nearly the same computational complexity. Moreover, since MS-JoCDWM-DL does not require iterative procedures, its inference complexity is significantly lower than that of MS-JoWM and MS-JoCDWM. \figref{time fig} compares the running time of several methods, where the time cost of SS-MMSE is relatively small and thus omitted. All experiments are conducted on a CPU platform (Intel\textregistered~Xeon\textregistered~W-2150B @ 3.00\,GHz). As observed, MS-JoCDWM-DL achieves significantly lower running time due to its non-iterative nature, even outperforming the single-satellite baseline SS-WM. It is worth noting that running time is influenced by various factors such as low-level implementation efficiency and hardware conditions, and should therefore be seen as indicative rather than absolute.

\section{Conclusion}\label{conclusion sec}
This paper investigated the distributed PD for multi-satellite cooperative transmission. We first conducted a detailed analysis of the transceiver model, examining the effects of delay and Doppler compensation errors and emphasizing the independence of inter-satellite interference signals. Then, we formulated the WSR problem that considers both sCSI and compensation errors. While similar problems are often recast as a WMMSE problem, we demonstrated that such problem is not equivalent to the considered WSR problem. Accordingly, we proposed an equivalent CDWMMSE problem, investigated a low-complexity matrix decomposition method, and proposed a solution algorithm. 
To reduce the computational complexity, we further propose a model-driven DL approach that exploits the inherent TE of the precoding mapping, supported by the proposed novel network architecture and scaling-invariant loss function. Simulation results demonstrated the effectiveness and robustness of the proposed method under representative practical scenarios.

    \appendices
\vspace{-3mm}
    \section{Proof of Proposition \ref{no equivalence ppn}}\label{no equivalence proof}
     \vspace{-1mm}
    For given ${\bf W}$, it can be proved that the optimal receiver is
  \begin{align}
    \textstyle a^{\star}_k=\frac{\mathbb{E}\{{\bar {\bf h}}^T_{k}\}{\bf w}_{k}}{\left(\sum_{i\neq k}{\bf w}^H_{i}{\tilde {\boldsymbol{\Omega}}}_k{\bf w}_{i}+{\bf w}^H_{k}{\boldsymbol{\Omega}}_k{\bf w}_{k}\right) + \sigma^2_k}.
    \label{optimal receiver}
  \end{align}
Note that this optimal expression is independent of the value of ${\bf u}$. With the optimal receiver and given ${\bf W}$, the optimal weight is $u^{\star}_k=e_k^{-1}$. By successively substituting the optimal expressions of ${\bf a}^{\star}$ and ${\bf u}^{\star}$ into the objective function of problem \eqref{wmmse problem}, we obtain the function that depends only on ${\bf W}$, which is expressed as \eqref{mse function expression}. Although the last term in the equation is similar to the sum rate expression in \eqref{WSR problem}, it is identical to the sum rate expression in \eqref{WSR problem} only when $\mathbb{E}\{{\bar {\bf h}}^*_{k}\}\mathbb{E}\{{\bar {\bf h}}^T_{k}\} = \mathbb{E}\{{\bar {\bf h}}^*_{k}{\bar {\bf h}}^T_{k}\}, \forall s\in\mathcal{S}$, which is equivalent to $\mathbb{E}\{{\bar {\bf h}}^*_{s,k}{\bar {\bf h}}^T_{s,k}\} = \mathbb{E}\{{\bar {\bf h}}^*_{s,k}\}\mathbb{E}\{{\bar {\bf h}}^T_{s,k}\},\ \forall s\in\mathcal{S}, k\in\mathcal{K}$. This concludes the proof.
\begin{figure*}[t]
    \begin{align}\textstyle
        &    \textstyle f  = \sum_{k=1}^{K}\beta_k \!-\! \beta_k\log_2(e_k^{-1}) = \sum_{k=1}^{K}\beta_k \!-\! \beta_k\log_2\left(1+\frac{{\bf w}^H_{k}\mathbb{E}\{{\bar {\bf h}}^*_{k}\}\mathbb{E}\{{\bar {\bf h}}^T_{k}\}{\bf w}_{k}}{\sum_{i\neq k}{\bf w}^H_{i}{\tilde {\boldsymbol{\Omega}}}_k{\bf w}_{i} + {\bf w}^H_{k}{\boldsymbol{\Omega}}_k{\bf w}_{k}-{\bf w}^H_{k}\mathbb{E}\{{\bar {\bf h}}^*_{k}\}\mathbb{E}\{{\bar {\bf h}}^T_{k}\}{\bf w}_{k}+ \sigma^2_k}\right).
        \label{mse function expression}
    \end{align}
    \vspace{-5mm} 
    \hrulefill  
\end{figure*}

\vspace{-3mm}
    \section{Proof of Proposition \ref{yes equivalence ppn}}\label{yes equivalence ppn proof}
     \vspace{-1mm}
    Given ${\bf W}$, the optimal virtual receive precoding vector is
    \begin{align}
        \textstyle{\bf a}^{\star}_k=\frac{{\bf Q}_{k}{\bf w}_{k}}{\left(\sum_{i\neq k}{\bf w}^H_{i}{\tilde {\boldsymbol{\Omega}}}_k{\bf w}_{i}+{\bf w}^H_{k}{\boldsymbol{\Omega}}_k{\bf w}_{k}\right) + \sigma^2_k}.
    \end{align}
    With the optimal receiver and given ${\bf W}$, the optimal weight is $u^{\star}_k={\tilde e}_k^{-1}$. By successively substituting the optimal expressions of ${\bf A}^{\star}$ and ${\bf u}^{\star}$ into the objective function of problem \eqref{modified wmmse problem}, we obtain the function that depends only on ${\bf W}$, which is expressed as follows:
        \begin{align}
        \textstyle
            {\tilde e}_k = \frac{\sum_{i\neq k}{\bf w}^H_{i}{\tilde {\boldsymbol{\Omega}}}_k{\bf w}_{i} + \sigma_k^2}{\left[{\left(\sum_{i\neq k}{\bf w}^H_{i}{\tilde {\boldsymbol{\Omega}}}_k{\bf w}_{i}+{\bf w}^H_{k}{\boldsymbol{\Omega}}_k{\bf w}_{k}\right) + \sigma^2_k}\right]},
        \end{align}
        and 
    \begin{align}
        &\textstyle\sum_{k=1}^{K}\beta_k\left(u_k{\tilde e}_k-\log_2(u_k)\right) = \sum_{k=1}^{K}\beta_k - \beta_k\log_2({\tilde e}_k^{-1})\notag\\
        &\textstyle= \sum_{k=1}^{K}\beta_k - \beta_k\log_2\left(1+\frac{{\bf w}^H_{k}{\boldsymbol{\Omega}}_k{\bf w}_{k}}{\sum_{i\neq k}{\bf w}^H_{i}{\tilde {\boldsymbol{\Omega}}}_k{\bf w}_{i} + \sigma_k^2}\right) \notag\\
        &\textstyle= \sum_{k=1}^{K}\beta_k - \beta_k{\bar R}_k.
    \end{align}
    Considering that $\beta_k,\forall k$ is a predefined non-negative value, minimizing the above function is equivalent to maximizing the objective function of optimization problem \eqref{WSR problem}. This concludes the proof.

    \vspace{-3mm}
    \section{Proof of Proposition \ref{optimal perUT ppn}}\label{optimal perUT ppn proof}
    \vspace{-1mm}
    According to the Stationarity Condition in the KKT conditions, the optimal ${\bf w}_k$ is given by the following expression:
    \begin{align}
        {\bf w}^{\star}_{k}(\eta_k,\mu_k) 
        = \eta_k({\boldsymbol{\Xi}}_k+\mu_k{\bf I})^{-1}{\bf d}_k,
    \end{align}
    where ${\bf d}_k=\beta_k u_k{\bf Q}^H_{k}{\bf a}_k$ and $\mu_k$ is the Lagrange multiplier associated with the power constraint. Consider the power constraint
    \begin{align}
        {\bf w}^{\star}_{k}(\mu_k) 
        &= \eta_k(\mu_k)({\boldsymbol{\Xi}}_k+\mu_k{\bf I})^{-1}{\bf d}_k, \\
        \eta_k(\mu_k) &\textstyle= \sqrt{\frac{P_k}{\|({\boldsymbol{\Xi}}_k+\mu_k{\bf I})^{-1}{\bf d}_k\|^2_2}}.
    \end{align}
    The cost function related to $\mu_k$ is 
    \begin{align}
        \textstyle f(\mu_k)= \frac{{\bf w}_k^H{\boldsymbol{\Xi}}_k{\bf w}_k}{\eta_k^2} - \frac{{\bf d}^H_k{\bf w}_k}{\eta_k}- \frac{{\bf w}^H_k{\bf d}_k}{\eta_k} + \frac{{\hat u}_k}{\eta_k^2},
    \end{align}
    where ${\hat u}_k = \beta_k u_k{\bf a}^H_k{\bf a}_k\sigma^2_k$. By deriving the gradient of this function with respect to $\mu_k$ based on total differentiation, we obtain $\mu_k = \frac{\beta_k u_k{\bf a}^H_k{\bf a}_k\sigma^2_k}{P_k}$ as one of the optimal solutions when the gradient is zero. This concludes the proof.

\ifCLASSOPTIONcaptionsoff
\newpage
\fi
 \vspace{-3mm}
\bibliographystyle{IEEEtran}
\bibliography{reference}

\begin{thebibliography}{10}
\providecommand{\url}[1]{#1}
\csname url@samestyle\endcsname
\providecommand{\newblock}{\relax}
\providecommand{\bibinfo}[2]{#2}
\providecommand{\BIBentrySTDinterwordspacing}{\spaceskip=0pt\relax}
\providecommand{\BIBentryALTinterwordstretchfactor}{4}
\providecommand{\BIBentryALTinterwordspacing}{\spaceskip=\fontdimen2\font plus
\BIBentryALTinterwordstretchfactor\fontdimen3\font minus \fontdimen4\font\relax}
\providecommand{\BIBforeignlanguage}[2]{{%
\expandafter\ifx\csname l@#1\endcsname\relax
\typeout{** WARNING: IEEEtran.bst: No hyphenation pattern has been}%
\typeout{** loaded for the language `#1'. Using the pattern for}%
\typeout{** the default language instead.}%
\else
\language=\csname l@#1\endcsname
\fi
#2}}
\providecommand{\BIBdecl}{\relax}
\BIBdecl
\renewcommand{\BIBentryALTinterwordstretchfactor}{4}

\bibitem{Ericsson2024}
\BIBentryALTinterwordspacing
{Ericsson}, ``{Ericsson joins MSSA to advance a non-terrestrial network ecosystem},'' Sep 2024, accessed: 2024-12-06. [Online]. Available: \url{https://www.ericsson.com/en/news/2024/9/ericsson-joins-mssa-to-advance-a-non-terrestrial-network-ecosystem}
\BIBentrySTDinterwordspacing

\bibitem{10820534}
K.~Ntontin, E.~Lagunas, J.~Querol, J.~u. Rehman \emph{et~al.}, ``A vision, survey, and roadmap toward space communications in the {6G} and beyond era,'' \emph{Proc. IEEE}, pp. 1--37, 2025.

\bibitem{3GPP_TR_38_811}
3GPP, ``Tr 38.811 v15.4.0: Study on new radio ({NR}) to support non-terrestrial networks,'' 3GPP, Tech. Rep. TR 38.811 V15.4.0, Sep. 2020.

\bibitem{3gpp_tr_38_821}
------, ``Tr 38.821 v16.2.0: Solutions for {NR} to support non-terrestrial networks ({NTN}),'' 3GPP, Tech. Rep. TR 38.821 V16.2.0, Mar. 2023.

\bibitem{9852737}
H.~Al-Hraishawi, H.~Chougrani, S.~Kisseleff, E.~Lagunas \emph{et~al.}, ``A survey on nongeostationary satellite systems: The communication perspective,'' \emph{IEEE Commun. Surv. Tutor.}, vol.~25, no.~1, pp. 101--132, First Quarter 2023.

\bibitem{DeFilippo2024CellFree}
B.~D. Filippo, R.~Campana, A.~Guidotti, C.~Amatetti \emph{et~al.}, ``Cell-free {MIMO} in {6G} {NTN} with {AI}-predicted {CSI},'' in \emph{Proc. IEEE 25th Int. Workshop Signal Process. Adv. Wireless Commun. (SPAWC)}, Sep. 2024, pp. 631--635.

\bibitem{8353925}
M.~{\'A}. V{\'a}zquez, M.~R.~B. Shankar, C.~I. Kourogiorgas, P.-D. Arapoglou \emph{et~al.}, ``Precoding, scheduling, and link adaptation in mobile interactive multibeam satellite systems,'' \emph{IEEE J. Sel. Areas Commun.}, vol.~36, no.~5, pp. 971--980, 2018.

\bibitem{8629918}
X.~Zhang, J.~Wang, C.~Jiang, C.~Yan \emph{et~al.}, ``Robust beamforming for multibeam satellite communication in the face of phase perturbations,'' \emph{IEEE Trans. Veh. Technol.}, vol.~68, no.~3, pp. 3043--3047, 2019.

\bibitem{wang2021resource}
W.~Wang, L.~Gao, R.~Ding, J.~Lei \emph{et~al.}, ``Resource efficiency optimization for robust beamforming in multi-beam satellite communications,'' \emph{IEEE Trans. Veh. Technol.}, vol.~70, no.~7, pp. 6958--6968, 2021.

\bibitem{you2020massive}
L.~You, K.-X. Li, J.~Wang, X.~Gao \emph{et~al.}, ``Massive {MIMO} transmission for {LEO} satellite communications,'' \emph{IEEE J. Sel. Areas Commun.}, vol.~38, no.~8, pp. 1851--1865, 2020.

\bibitem{10437228}
S.~Wu, Y.~Wang, G.~Sun, L.~You \emph{et~al.}, ``Energy and computational efficient precoding for {LEO} satellite communications,'' in \emph{Proc. IEEE Glob. Commun. Conf. (GLOBECOM)}, Dec. 2023, pp. 1872--1877.

\bibitem{9815078}
Y.~Liu, Y.~Wang, J.~Wang, L.~You \emph{et~al.}, ``Robust downlink precoding for {LEO} satellite systems with per-antenna power constraints,'' \emph{IEEE Trans. Veh. Technol.}, vol.~71, no.~10, pp. 10\,694--10\,711, Oct. 2022.

\bibitem{10336551}
M.~Alsenwi, E.~Lagunas, and S.~Chatzinotas, ``Robust beamforming for massive {MIMO} {LEO} satellite communications: A risk-aware learning framework,'' \emph{IEEE Trans. Veh. Technol.}, vol.~73, no.~5, pp. 6560--6571, May 2024.

\bibitem{10550141}
M.~Ying, X.~Chen, Q.~Qi, and W.~Gerstacker, ``Deep learning-based joint channel prediction and multibeam precoding for {LEO} satellite internet of things,'' \emph{IEEE Trans. Wireless Commun.}, vol.~23, no.~10, pp. 13\,946--13\,960, Oct. 2024.

\bibitem{9939157}
M.~Y. Abdelsadek, G.~K. Kurt, and H.~Yanikomeroglu, ``Distributed massive {MIMO} for {LEO} satellite networks,'' \emph{IEEE Open J. Commun. Soc.}, vol.~3, pp. 2162--2177, Nov. 2022.

\bibitem{10061620}
M.~Y. Abdelsadek, G.~Karabulut-Kurt, H.~Yanikomeroglu, P.~Hu \emph{et~al.}, ``Broadband connectivity for handheld devices via {LEO} satellites: Is distributed massive {MIMO} the answer?'' \emph{IEEE Open J. Commun. Soc.}, vol.~4, pp. 713--726, Mar. 2023.

\bibitem{10380500}
X.~Zhang, S.~Sun, M.~Tao, Q.~Huang \emph{et~al.}, ``Multi-satellite cooperative networks: Joint hybrid beamforming and user scheduling design,'' \emph{IEEE Trans. Wireless Commun.}, pp. 1--1, Jul. 2024.

\bibitem{10440321}
Z.~Xiang, X.~Gao, K.-X. Li, and X.-G. Xia, ``Massive {MIMO} downlink transmission for multiple {LEO} satellite communication,'' \emph{IEEE Trans. Commun.}, vol.~72, no.~6, pp. 3352--3364, Jun. 2024.

\bibitem{ha2024UCB}
V.~N. Ha, D.~H.~N. Nguyen, J.~C.-M. Duncan, J.~L. Gonzalez-Rios \emph{et~al.}, ``User-centric beam selection and precoding design for coordinated multiple-satellite systems,'' in \emph{Proc. IEEE 35th Int. Symp. Personal, Indoor and Mobile Radio Commun. (PIMRC)}, Sept. 2024, pp. 1--6.

\bibitem{10596023}
X.~Chen and Z.~Luo, ``Asynchronous interference mitigation for {LEO} multi-satellite cooperative systems,'' \emph{IEEE Trans. Wireless Commun.}, vol.~23, no.~10, pp. 14\,956--14\,971, Oct. 2024.

\bibitem{10615897}
------, ``Cooperative {WMMSE} precoding for asynchronous {LEO} multi-satellite communications,'' in \emph{Proc. IEEE Int. Conf. Commun. Workshops (ICC Workshops)}, Jun. 2024, pp. 1691--1696.

\bibitem{wu2025distributed}
S.~Wu, Y.~Wang, G.~Sun, W.~Wang \emph{et~al.}, ``Distributed beamforming for multiple {LEO} satellites with imperfect delay and {Doppler} compensations: Modeling and rate analysis,'' \emph{IEEE Trans. Veh. Technol.}, pp. 1--6, May 2025, {Early Access}.

\bibitem{8840846}
L.~Bai, C.-X. Wang, G.~Goussetis, S.~Wu \emph{et~al.}, ``Channel modeling for satellite communication channels at {Q}-band in high latitude,'' \emph{IEEE Access}, vol.~7, pp. 137\,691--137\,703, 2019.

\bibitem{8795582}
W.~Wang, Y.~Tong, L.~Li, A.-A. Lu \emph{et~al.}, ``Near optimal timing and frequency offset estimation for {5G} integrated {LEO} satellite communication system,'' \emph{IEEE Access}, vol.~7, pp. 113\,298--113\,310, 2019.

\bibitem{wu2023low}
S.~Wu, G.~Sun, Y.~Wang, L.~You \emph{et~al.}, ``Low-complexity user scheduling for {LEO} satellite communications,'' \emph{IET Commun.}, vol.~17, no.~12, pp. 1368--1383, May 2023.

\bibitem{yuan2023alternating}
M.~Yuan, H.~Wang, H.~Yin, and D.~He, ``Alternating optimization based hybrid transceiver designs for wideband millimeter-wave massive multiuser {MIMO}-{OFDM} systems,'' \emph{IEEE Trans. Wireless Commun.}, vol.~22, no.~12, pp. 9201--9217, Dec. 2023.

\bibitem{bjornson2014optimal}
E.~Bj{\"o}rnson, M.~Bengtsson, and B.~Ottersten, ``Optimal multiuser transmit beamforming: A difficult problem with a simple solution structure [lecture notes],'' \emph{{IEEE Signal Process. Mag.}}, vol.~31, no.~4, pp. 142--148, Jul. 2014.

\bibitem{SB-9931}
M.~Bengtsson and B.~Ottersten, ``Optimal downlink beamforming using semidefinite optimization,'' in \emph{Proc. 37th Annual Allerton Conference on Communication, Control, and Computing}, Sep. 1999, pp. 987--996, invited paper.

\bibitem{SB-0105}
------, ``Optimal and suboptimal transmit beamforming,'' in \emph{Handbook of Antennas in Wireless Communications}, L.~C. Godara, Ed.\hskip 1em plus 0.5em minus 0.4em\relax CRC Press, Aug. 2001.

\bibitem{christensen2008weighted}
S.~S. Christensen, R.~Agarwal, E.~De~Carvalho, and J.~M. Cioffi, ``Weighted sum-rate maximization using weighted {MMSE} for {MIMO-BC} beamforming design,'' \emph{IEEE Trans. Wireless Commun.}, vol.~7, no.~12, pp. 4792--4799, Dec. 2008.

\bibitem{shi2011iteratively}
Q.~Shi, M.~Razaviyayn, Z.-Q. Luo, and C.~He, ``An iteratively weighted {MMSE} approach to distributed sum-utility maximization for a {MIMO} interfering broadcast channel,'' \emph{IEEE Trans. Signal Process.}, vol.~59, no.~9, pp. 4331--4340, Sept. 2011.

\bibitem{ASTSpaceMobile2023}
{AST SpaceMobile}, ``{AST SpaceMobile achieves space-based 5G cellular broadband connectivity from everyday smartphones, another historic world first},'' Sep 2023, accessed: 2024-12-06. \url{https://ast-science.com/2023/09/19/ast-spacemobile-achieves-space-based-5g-cellular-broadband-connectivity-from-everyday-smartphones-another-historic-world-first}.

\bibitem{wang2024towards}
Y.~Wang, H.~Hou, X.~Yi, W.~Wang \emph{et~al.}, ``Towards unified {AI} models for {MU-MIMO} communications: A tensor equivariance framework,'' arXiv preprint arXiv:2406.09022, Jun. 2024, available: \url{https://arxiv.org/abs/2406.09022}.

\bibitem{zaheer2017deep}
M.~Zaheer, S.~Kottur, S.~Ravanbakhsh, B.~Poczos \emph{et~al.}, ``Deep sets,'' \emph{Neural Inf. Proces. Syst. (NeurIPS)}, Long Beach, CA, United states, Dec. 2017.

\bibitem{9298921}
K.~Pratik, B.~D. Rao, and M.~Welling, ``{RE-MIMO}: Recurrent and permutation equivariant neural {MIMO} detection,'' \emph{IEEE Trans. Signal Process.}, vol.~69, pp. 459--473, Jan. 2021.

\bibitem{10584439}
Y.~Wang, H.~Hou, W.~Wang, X.~Yi \emph{et~al.}, ``Soft demodulator for symbol-level precoding in coded multiuser {MISO} systems,'' \emph{IEEE Trans. Wireless Commun.}, vol.~23, no.~10, pp. 14\,819--14\,835, Oct. 2024.

\bibitem{9844981}
J.~Kim, H.~Lee, S.-E. Hong, and S.-H. Park, ``A bipartite graph neural network approach for scalable beamforming optimization,'' \emph{IEEE Trans. Wireless Commun.}, vol.~22, no.~1, pp. 333--347, Jan. 2023.

\bibitem{bronstein2021geometric}
M.~M. Bronstein, J.~Bruna, T.~Cohen, and P.~Veli{\v{c}}kovi{\'c}, ``Geometric deep learning: Grids, groups, graphs, geodesics, and gauges,'' \emph{arXiv preprint arXiv:2104.13478}, Apr. 2021, available: \url{https://arxiv.org/abs/2104.13478}.

\bibitem{vaswani2017attention}
A.~Vaswani, N.~Shazeer, N.~Parmar, J.~Uszkoreit \emph{et~al.}, ``Attention is all you need,'' in \emph{Adv. Neural Inf. Process. Syst. (NeurIPS)}, vol.~30, 2017.

\bibitem{zhu2025transformersnormalization}
J.~Zhu, X.~Chen, K.~He, Y.~LeCun \emph{et~al.}, ``Transformers without normalization,'' in \emph{Proc. IEEE/CVF Conf. Comput. Vis. Pattern Recognit. (CVPR)}, Jun. 2025.

\bibitem{yun2019transformers}
C.~Yun, S.~Bhojanapalli, A.~S. Rawat, S.~J. Reddi \emph{et~al.}, ``Are transformers universal approximators of sequence-to-sequence functions?'' arXiv preprint arXiv:1912.10077, Dec. 2019, available: \url{https://arxiv.org/abs/1912.10077}.

\bibitem{huang2017densely}
G.~Huang, Z.~Liu, L.~V.~D. Maaten, and K.~Q. Weinberger, ``Densely connected convolutional networks,'' in \emph{Proc. IEEE Conf. Comput. Vis. Pattern Recognit. (CVPR)}, Jul. 2017, pp. 4700--4708.

\bibitem{hartford2018deep}
J.~Hartford, D.~Graham, K.~Leyton-Brown, and S.~Ravanbakhsh, ``Deep models of interactions across sets,'' in \emph{Int. Conf. Mach. Learn. (ICML)}, vol.~5, Stockholm, Sweden, Jul. 2018, pp. 3050--3061.

\bibitem{lee2019set}
J.~Lee, Y.~Lee, J.~Kim, A.~Kosiorek \emph{et~al.}, ``Set transformer: {A} framework for attention-based permutation-invariant neural networks,'' in \emph{Int. Conf. Mach. Learn. (ICML)}, vol.~97, Jun. 2019, pp. 3744--3753.

\bibitem{6902008}
F.~Burkhardt, S.~Jaeckel, E.~Eberlein, and R.~Prieto-Cerdeira, ``{QuaDRiGa}: A {MIMO} channel model for land mobile satellite,'' in \emph{Proc. 8th European Conf. Antennas Propag. (EuCAP 2014)}, Apr. 2014, pp. 1274--1278.

\bibitem{9815679}
S.~Jaeckel, L.~Raschkowski, and L.~Thieley, ``A {5G-NR} satellite extension for the {QuaDRiGa} channel model,'' in \emph{Proc. Joint Eur. Conf. Netw. Commun. \& 6G Summit (EuCNC/6G Summit)}, 2022, pp. 142--147.

\bibitem{QuaDRiGa2023}
\BIBentryALTinterwordspacing
{Fraunhofer Heinrich Hertz Institute}, \emph{Quasi Deterministic Radio Channel Generator: User Manual and Documentation}, v2.8.1~ed., Einsteinufer 37, 10587 Berlin, Germany, Dec. 2023, available: \url{https://github.com/fraunhoferhhi/QuaDRiGa}. [Online]. Available: \url{http://www.quadriga-channel-model.de}
\BIBentrySTDinterwordspacing

\bibitem{SpaceX_Gen2_2021}
``Federal communications commission; amendment to pending application for the {SpaceX} {Gen2} {NGSO} satellite system,'' FCC, Washington, D.C., Tech. Rep. File No. SAT-AMD-2021, August 2021, available: https://fcc.report/IBFS/SAT-AMD-20210818-00105/12943361.pdf.

\bibitem{9998075}
L.~Yu, J.~Wan, K.~Zhang, F.~Teng \emph{et~al.}, ``Spaceborne multibeam phased array antennas for satellite communications,'' \emph{IEEE Aerosp. Electron. Syst. Mag.}, vol.~38, no.~3, pp. 28--47, Mar. 2023.

\end{thebibliography}
	
\end{document}